\documentclass{pinchcr}   % Comment the above and uncomment this
\title{Quantum optics with nitrogen-vacancy centers in diamond}

\author{Yiwen Chu and Mikhail D. Lukin}
\affiliation{Department of Physics, Harvard University, Cambridge,
MA 02138 USA}
\authors{2}

\begin{document}

%%% mathdefs.tex
%

\newcommand{\figSize}{0.5}
\newcommand{\ExState}{|\mathrm{E}_\mathrm{x}\rangle}
\newcommand{\EyState}{|\mathrm{E}_\mathrm{y}\rangle}
\newcommand{\Ex}{|0\rangle\leftrightarrow|\mathrm{E}_\mathrm{x}\rangle}
\newcommand{\Ey}{|0\rangle\leftrightarrow|\mathrm{E}_\mathrm{y}\rangle}
\newcommand{\gPar}{g^{(2)}_{||}}
\newcommand{\gPerp}{g^{(2)}_{\bot}}
\newcommand{\gPSB}{g^{(2)}_{\mathrm{PSB}}}

% Taken from Adam Lupu-Sax ....................................................
%%% derivatives
\def\rp#1{\left(#1\right)}
\def\kp#1{\left\{#1\right\}}
\def\sqp#1{\left[#1\right]}
\def\bp#1{\left|#1\right|}

\def\ketbra#1#2{|{#1}\rangle\langle{#2}|}

\newcommand{\deriv}[2]{\frac{d#1}{d#2}}
\newcommand{\derivc}[3]{\left. \frac{d#1}{d#2}\right|_{#3}}
\newcommand{\pd}[2]{\frac{\partial #1}{\partial #2}}
\newcommand{\pdc}[3]{\left. \frac{\partial #1}{\partial #2}\right|_{#3}}

%%% Dirac notation
\newcommand{\bra}[1]{\left\langle #1\right|}
\newcommand{\ket}[1]{\left|#1\right\rangle}
\newcommand{\braket}[2]{\left\langle #1 \left|#2\right.\right\rangle}
\newcommand{\braOket}[3]{\left\langle #1\left|#2\right|#3\right\rangle}

% calligraphic letters in math.
\def\cal#1{\mathcal{#1}}

% Taken from M Haggerty ......................................................
\def\avg#1{\left< #1 \right>}
\def\abs#1{\left| #1 \right|}
\def\recip#1{\frac{1}{#1}}
\def\vhat#1{\hat{{\bf #1}}}
\def\smallfrac#1#2{{\textstyle\frac{#1}{#2}}}
\def\smallrecip#1{\smallfrac{1}{#1}}

% SPSmith's definitions ......................................................
\def\spshalf{{1\over{2}}}
\def\Orabi{\Omega_{\rm rabi}}
\def\btt#1{{\tt$\backslash$#1}}

% My own .....................................................................

%%% Equations
\def\schrod{Schroedinger's Equation}
\def\helm{Helmholtz Equation}

%%% Equation environments
\def\be{\begin{equation}}
\def\ee{\end{equation}}
\def\bea{\begin{eqnarray}}
\def\eea{\end{eqnarray}}
\def\bean{\begin{mathletters}\begin{eqnarray}}
\def\eean{\end{eqnarray}\end{mathletters}}

%%% macros for Feb 2000 PRL
\newcommand{\tbox}[1]{\mbox{\tiny #1}}
\newcommand{\half}{\mbox{\small $\frac{1}{2}$}}
\newcommand{\pit}{\mbox{\small $\frac{\pi}{2}$}}
\newcommand{\sfrac}[1]{\mbox{\small $\frac{1}{#1}$}}
\newcommand{\mbf}[1]{{\mathbf #1}}
% hack to get APS's \text style to work ok:
\def\text{\tbox}

\newcommand{\mV}{{\mathsf{V}}}
\newcommand{\mL}{{\mathsf{L}}}
\newcommand{\mA}{{\mathsf{A}}}
\newcommand{\lB}{\lambda_{\tbox{B}}}  % de Broglie
\newcommand{\ofr}{{(\mbf{r})}}       % (r vec)
\def\ofkr{(k;\mbf{r})}      % (k;r vec)
\def\ofks{(k;\mbf{s})}      % (k;s vec)
\newcommand{\ofs}{{(\mbf{s})}}       % (s vec)
\def\xt{\mbf{x}^{\tbox T}}    % x^T vec

\def\ce{\tilde{C}_{\tbox E}}    % C_E
\def\cew{\tilde{C}_{\tbox E}(\omega)}    % C_E(w)
\def\ceqmw{\tilde{C}^{\tbox{qm}}_{\tbox E}(\omega)}  % C^qm_E(w)
\def\cewqm{\tilde{C}^{\tbox{qm}}_{\tbox E}}  % C^qm_E(w) no omega
\def\ceqm{C^{\tbox{qm}}_{\tbox E}}  % C^qm_E no tilde
\def\cw{\tilde{C}(\omega)}    % C(w)
\def\cfw{\tilde{C}_{\cal F}(\omega)}    % C_F(w)

\def\tcl{\tau_{\tbox{cl}}}    % tau cl
\def\tcol{\tau_{\tbox{col}}}    % tau col
\def\terg{t_{\tbox{erg}}}    % t_erg
\def\tbl{\tau_{\tbox{bl}}}    % tau bl
\def\theis{t_{\tbox{H}}}    % t_Heis

\def\area{\mathsf{A}_D}      % piston effective area A_D 
\def\ve{\nu_{\tbox{E}}}      % nu_E, noise intensity
\def\vewna{\nu_E^{\tbox{WNA}}}    % nu_E for WNA

\def\dxcqm{\delta x^{\tbox{qm}}_{\tbox c}}  % x to mix levels.

%% operators
\newcommand{\rop}{\hat{\mbf{r}}}  % vector valued operators
\newcommand{\pop}{\hat{\mbf{p}}}

%%% Integrals
\newcommand{\sint}{\oint \! d\mbf{s} \,} % surface int
\def\gint{\oint_\Gamma \!\! d\mbf{s} \,} % surface int over Gamma.
\newcommand{\lint}{\oint \! ds \,}  % d=2
\def\infint{\int_{-\infty}^{\infty} \!\!}  % infinite integral
\def\dn{\partial_n}        % d_n
\def\aswapb{a^*\!{\leftrightarrow}b}    % a* <-> b
\def\eps{\varepsilon}        % eps

%%% Dissipation.
\def\dhdxt{\partial {\cal H} / \partial x}
\def\dhdx{\pd{\cal H}{x}}
\def\dhdxnm{\left( \pd{\cal H}{x} \right)_{\!nm}}
\def\dhdxnmsq{\left| \left( \pd{\cal H}{x} \right)_{\!nm} \right| ^2}

%%% vergini
\def\bcs{\stackrel{\tbox{BCs}}{\longrightarrow}}  % apply BCs

%%% DIEL atom project.
\def\wx{\omega_x}
\def\wy{\omega_y}
\newcommand{\ofro}{({\bf r_0})}
\def\Eb{E_{\rm blue,rms}}
\def\Er{E_{\rm red,rms}}
\def\Es2{E_{0,{\rm sat}}^2}
\def\sb{s_{\rm blue}}
\def\sr{s_{\rm red}}

%%% text usefuls
\def\ie{{\it i.e.\ }}
\def\eg{{\it e.g.\ }}
\newcommand{\etal}{{\it et al.\ }}
\newcommand{\ibid}{{\it ibid.\ }}

%%% tables spaces.
\def\gap{\hspace{0.2in}}

%%%%%%%%%%%%%%%%%%%%
%%% mathletters code (from http://www.grad.uiuc.edu/thesis/latexcode.html )
% Modified Alex Barnett 00/9/28 to handle non-arabic \thechapter output
% (ie, now works in Appendices)
% Fails to give correct references in eqnarray (offset by one).
% Still inserts small extra space after equation - unknown reason.
%
%    Original code:
%\newcounter{eqletter}
%\def\mathletters{%
%\setcounter{eqletter}{0}%
%\addtocounter{equation}{1}
%\edef\curreqno{\arabic{equation}}
%\edef\@currentlabel{\theequation}
%\def\theequation{%
%\addtocounter{eqletter}{1}\arabic{chapter}.\curreqno\alph{eqletter}%
%}%
%}
%\def\endmathletters{\setcounter{equation}{\curreqno}}

\newcounter{eqletter}
\def\mathletters{%
\setcounter{eqletter}{0}%
\addtocounter{equation}{1}
\edef\curreqno{\arabic{equation}}
\edef\@currentlabel{\theequation}
\def\theequation{%
\addtocounter{eqletter}{1}\thechapter.\curreqno\alph{eqletter}%
}%
}
\def\endmathletters{\setcounter{equation}{\curreqno}}

%...............................................QPC...................

\newcommand{\bk}{{\bf k}}
\def\kf{k_{\text F}}
\newcommand{\br}{{\bf r}}
\newcommand{\TL}{{\text{(L)}}}
\newcommand{\TR}{{\text{(R)}}}
\newcommand{\TLR}{{\text{L,R}}}
\newcommand{\VSD}{V_{\text{SD}}}
\newcommand{\GT}{\Gamma_{\text{T}}}
\newcommand{\DEL}{\mbox{\boldmath $\nabla$}}
\def\lf{\lambda_{\text F}}
\def\st{\sigma_{\text T}}
\def\stlr{\sigma_{\text T}^{\text{L$\rightarrow$R}}}
\def\strl{\sigma_{\text T}^{\text{R$\rightarrow$L}}}
\def\aeff{a_{\text{eff}}}
\def\aaeff{A_{\text{eff}}}
\def\gat{G_{\text{atom}}}
\newcommand{\LB}{Landauer-B\"{u}ttiker}

% .............. NOTES ..................
% To force linebreak when get overfull hbox from equation in paragraph
% test, use \linebreak (which makes it justify the line it broke).
% Contrast \\ or \newline which leave empty space.
%
%
 % my math definitions.
\newcommand{\figref}[1]{Fig.~{#1}}
\newcommand{\eqnref}[1]{Eq.~\eqref{#1}}
% first derivative
\newcommand{\dbydt}[2]{\ensuremath{\frac{d{#1}}{d{#2}}}}
%will print the first partial derivative
\newcommand {\doe}[2]{\ensuremath{\frac{\partial {#1}}{\partial {#2}}}}

\newcommand{\vect}[1]{\ensuremath{\mathbf{#1}}}

%will print arg.1 in bold, and with a vector sign and subscript arg.2,meant for greek symbols
\newcommand{\bvecsub}[2]{\ensuremath{\mbox{\boldmath${#1}$}_{#2}}}
%will print arg.1 in bold, meant for greek symbols, it works but amsmath package has boldsymbol command that also works
%\newcommand{\bvec}[1]{\ensuremath{\mbox{\boldmath${#1}$}}}
\newcommand{\bvec}[1]{\boldsymbol{#1}}
%dot product of two vectors
\newcommand{\dotpr}[2]{\ensuremath{\bvec{#1}\cdot \bvec{#2}}}
%will print the rotation operator
\newcommand{\rotop}{\ensuremath{\mathcal{D}}}
%will print magnitude symbol around quantity
\newcommand {\magsym}[1]{\ensuremath{\bigl|#1\bigr|}}
%will print 13C
\newcommand{\carb}{\ensuremath{^{13}\mbox{C }}}
%will print 14N
\newcommand{\fortn}{\ensuremath{^{14}\mbox{N }}}

\maketitle
%
%\dedication{This is an example of a dedication.}
%
%\preface

%Blah blah

%\medskip

%This is an example of a Preface which has been created with the \verb+\preface+ command, see
%Section~\ref{other prelims}.

%\acknowledgements

%This is an example of an Acknowledgements section which has been implemented with
%the \verb+\acknowledgements+ command.

\tableofcontents

\maintext

%&&&&&&&&&&&&&&&&&&&&&&&&&&&&&&&&&
%\part{The \LaTeX\ typesetting system}
%&&&&&&&&&&&&&&&&&&&&&&&&&&&&&&&&&
%
%\label{Part: LaTeX typesetting}

%\LaTeX\ is a typesetting system which is very suitable for
%producing scientific and mathematical documents of high
%typographical quality. The system
%is also suitable for producing all sorts of other documents, from
%simple letters to complete books. \LaTeX\ uses \TeX\ as its
%formatting engine.

%\LaTeX\ is available for most computers, from the PC and Mac to large
%UNIX and VMS systems.

%If you need to obtain any \LaTeX\ related material, see
%one of the Comprehensive \TeX\ Archive Network (\texttt{CTAN}) ftp archives. They
%can be found, for example, at the WWW address
%\texttt{http://www.tex.ac.uk}, where a large volume of styles and
%options can be downloaded.
%Further, this WWW address also provides detailed descriptions and
%additional links on how to install and run the basic
%\LaTeX\ software on your own computer.

\chapter{Quantum optics with nitrogen-vacancy centers in diamond}

\section{Background}

The study of classical mechanics, thermodynamics, and electromagnetism have brought on great technological advances over the course of history. We now live in an exciting era where quantum mechanics begins to join these other fields of physics in enabling new technologies with unprecedented capabilities. Theoretical and experimental efforts have moved toward harnessing the power of various different physical systems for applications in metrology, quantum information, and furthering our understanding of quantum mechanics itself. 

Some of the most successful of applications of quantum mechanical systems to date have been in the field of metrology, or the accurate sensing and measurement of forces, fields, frequencies, etc.  \shortcite{ChouPRL2010,NicholsonPRL2012,Benko2012,KesslerNatPhot2012}. 
%One important example of such a quantum technology is the optical clock, which provides high accuracy standards for frequency and time based on the fundamental stability of transition frequencies between quantum states of neutral atoms or trapped ions \shortcite{ChouPRL2010,NicholsonPRL2012}. One challenge of implementing such optical clocks is lies in reducing the uncertainty of the clock transition frequency caused by external effects. To this end, for example, different atomic species and level structures have been extensively studied to find transitions between energy levels that are particularly insensitive to external electromagnetic fields. In addition, the atoms are trapped and cooled using ion traps or optical lattices to reduce the effect of atomic motion. The other major challenge is developing techniques for probing the transition frequency with high enough resolution and stability. This has led to the development of, for example, high quality optical cavities for laser stabilization and the use of frequency combs for analyzing and comparing the accuracy of different clocks \shortcite{Benko2012,KesslerNatPhot2012}. 
At the same time, the field of quantum information and communications has become the focus of much effort in recent years. There has been a large number of theoretical proposals for using quantum systems for efficient computation, simulation of complex systems, secure dissemination of information, etc. \shortcite{Shor1994,Kimble:Nature2008,LloydScience1996}. However, the practical implementation of these proposals in real physical systems remains an open challenge. A variety of different avenues are being actively explored, ranging from photons, neutral atoms, ions, impurities and structures in solid state systems, to more exotic materials such as electrons on the surface of liquid helium \shortcite{MonroeScience2013,AwschalomScience2013,ClarkeNature2008,SternScience2013,PlatzmanScience1999}. 
%To get a sense of both the exciting developments and difficult challenges of this field, one can look at the example of superconducting quantum bits (qubits). This class of qubits consists of devices constructed from superconducting circuit elements such as Josephson junctions and engineered to satisfy requirements such as energy level anharmonicity, addressability,  and immunity to external perturbations \shortcite{ClarkeNature2008}. When these qubits are integrated with resonator and waveguide structures, they can be initialized, manipulated, and read out with very high fidelity through microwave fields \shortcite{RistePRL2012,MalletNatPhys2009}. In addition, the strong coupling achievable between superconducting qubits and microwave resonators have led to demonstrations of multi-qubit gates and operations that pave the way towards scalable architectures for quantum computation \shortcite{ReedNature2012,SteffenNature2013}. As complex mesoscopic systems, however, these qubits can be highly sensitive to their environment, which leads to loss of quantum information through decoherence. Tremendous progress has been made to isolate these systems from the environment, with qubit relaxation and decoherences times increasing from nanoseconds to nearly 0.1 ms in a little over a decade \shortcite{Nakamura99,Paik11,RigettiPRB2012}. However, fully understanding the underlying mechanisms for decoherence in these systems will require further studies in the fields of material science, surface chemistry, and the physics of superconductors \shortcite{MartinisPRL2005,KochPRL2007}. 
A common theme for the implementation of quantum technologies involves addressing the seemingly contradictory need for controllability and isolation from external effects. Undesirable effects of the environment must be minimized, while at the same time techniques and tools must be developed that enable us to interact with the system in a controllable and well-defined manner. These lectures address several aspects of this theme with regards to a particularly promising candidate for developing applications in both metrology and quantum information: the nitrogen-vacancy (NV) center in diamond. In particular, we explore how the NV center can be investigated, controlled, and efficiently coupled with optical photons, and tackle the challenges of controlling the optical properties of these emitters inside a complex solid-state environment. Some recent reviews on NV centers can be found in \shortcite{Wrachtrup:Review2006,DohertyReview20131,DohertyReview2012}.
%A tremendous amount progress has been made in using the NV center for local storage and manipulation of quantum information. The work presented here not only demonstrates its ability to also act as an interface with light, which will allow for the dissemination and . 

As with other solid state systems such as quantum dots, rare earth dopants in inorganic crystals, and superconducting qubits, NV centers present the attractive feature of eliminating the need for trapping and cooling, as is usually required for atomic systems. In addition to greatly reducing the complexity of experiments, solid state systems offer the possibility of straightforward integration with fabricated structures and development of scalable quantum devices. The issue of scalability, however, is not simply a matter of being able to create a large number of physical devices in a small volume. In order to truly move beyond proof-of-principle experiments, the properties of many qubits and devices must all be well controlled in a reproducible, reliable, and efficient manner. In other words, while it is important to demonstrate that a certain quantum operation, coherence time, metrological sensitivity, etc. is possible in one particular device with favorable characteristics, it must also be possible to reproduce these characteristics for a large number of systems. 
%
%\chapter{Optical properties of the NV center}
%\label{SpectroscopyChapter}

%\section{Introduction}

%Optical spectroscopy is a powerful tool for determining the structure and behavior of quantum systems, and has historically been the main experimental method contributing to the understanding of atoms, ions, and molecules. There are now well-established methods in the field of atomic, optical and molecular physics for trapping, cooling, and manipulating the internal states of these systems with light. More recently, however, there has been significant interest in discovering and exploring solid state systems that exhibit atom-like optical properties for the goal of building scalable, compact devices. Great progress has been made in understanding and utilizing the optical properties of, for example, self-assembled quantum dots, dye molecules embedded in organic matrix materials, rare earth ions in doped crystals, and impurities in insulators and semiconductors \shortcite{Gupta2001Science,RezusPRL2012,RiedmattenNature2008,FalkNatCom2013}. 

Diamond is a particularly promising host material for optically active quantum emitters due to its wide bandgap (5.5 eV) and large transparency window. It supports a huge variety of impurities and defects that give rise to optical bands at energies throughout the infrared and visible spectrum \shortcite{Zaitsev10}. The NV center is unique in that it combines well-defined optical transitions with long-lived spin sub-levels in the ground state, much like alkali atoms with transitions involving hyperfine sublevels. The present lectures focus on quantum optical experiments with NV centers. In recent years, a comprehensive theoretical model has been developed to describe the spin and optical properties of the NV center, which has been reviewed in \shortcite{MazeNJP2011,DohertyReview2012,DohertyArxiv2010}. We begin by summarizing this theoretical framework, which allows us to understand the underlying physics of the subsequent experimental results. We then describe the experimental techniques that are used to investigate and address the optical properties of NV centers. As part of this description, we will encounter some general characteristics of the system that will be come important issues to be addressed later, such as the Debye-Waller factor and the sensitivity of the excited state to the local environment. Based on our theoretical understanding and spectroscopic results, we show that NV centers offer a rich set of possibilities for experiments demonstrating coherent optical control of the system, several of which are summarized in the remainder of the chapter.

\section{Level structure and polarization properties of the NV center}
\label{OpticalProperties}

As is the case in any system, the electronic structure of the NV center can be analyzed by considering its symmetry, in addition to the various interactions that are present. Our description below starts with electronic states that obey the $C_{3v}$ symmetry of the NV center. We then consider the effect of various interactions one by one, starting from the the Coulomb interaction, which sets the energy scale of the optical transitions. Spin-orbit and spin-spin interactions act to lift the degeneracy of the different states in the ground and excited state manifolds. Finally, we consider the effect of external perturbations such as electric and magnetic fields that change the nature and energies of the unperturbed states. More detailed analysis of many of the concepts presented in this section can be found in \shortcite{MazeNJP2011}.

\subsection{Electronic states of the NV center}
\begin{figure}
\centering
\includegraphics[width=0.9\textwidth]{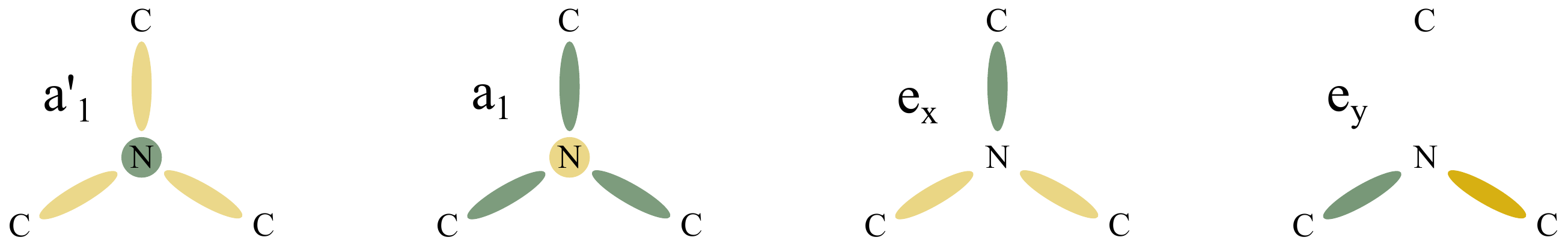}
\caption{Single electron orbitals of the NV center. The nitrogen and three carbons surrounding the vacancy are shown, and the perspective is along the NV axis. The color scale roughly represents the sign and occupation of each orbital.}
\label{f:orbitals}
\end{figure}

The negatively-charged NV center has six electrons, five of which are from the nitrogen
and the three carbons surrounding the vacancy. They occupy the
orbitals states $a'_1, a_1,e_x, e_y$, which are combinations of the four dangling bonds surrounding the vacancy that satisfy the symmetry imposed by the
nuclear potential. They can be written as 
\bea
a'_1&=&\alpha \frac{\sigma_1+\sigma_2+\sigma_3}{3}+\beta \sigma_n,\\
a_1&=&\beta \frac{\sigma_1+\sigma_2+\sigma_3}{3}+\alpha \sigma_n,\\
e_x&=&\frac{2\sigma_1-\sigma_2-\sigma_3}{\sqrt{6}},\\
e_y&=&\frac{\sigma_2-\sigma_3}{\sqrt{2}}
\eea
where $\sigma_1$, $\sigma_2$, $\sigma_3$ and $\sigma_n$ are the dangling bonds from the three carbons and the nitrogen, respectively.  These orbitals transform as the irreducible
representations of the $C_{3v}$ group and have been extensively used
by many authors \shortcite{Loubser:RPP1978,Goss:PRL1996,Lenef:PRB1996,Gali:PRB2008}.  They are schematically illustrated in Figure \ref{f:orbitals} to give an idea of their symmetries. From these pictorial representations, one can deduce that, for example, $\bra{e_y}\hat{y}\cdot\vec{r}\ket{e_x}=\bra{e_x}\hat{x}\cdot\vec{r}\ket{e_x}=-\bra{e_y}\hat{x}\cdot\vec{r}\ket{e_y}\neq0$. This means that the $\ket{e_x}$ and $\ket{e_y}$ states have permanent electric dipole moments, which will become important later for understanding the electric field sensitivity of the NV center's excited states and spectral diffusion of the optical transitions. In addition, one also sees that $\bra{a}\hat{x}\cdot\vec{r}\ket{e_x}=\bra{a}\hat{y}\cdot\vec{r}\ket{e_y}\neq0$, which gives rise to the transition dipole moments between the ground and excited states. These properties will be discussed in more detail in the following sections. 

\begin{figure}
\centering
\includegraphics[width=0.9\textwidth]{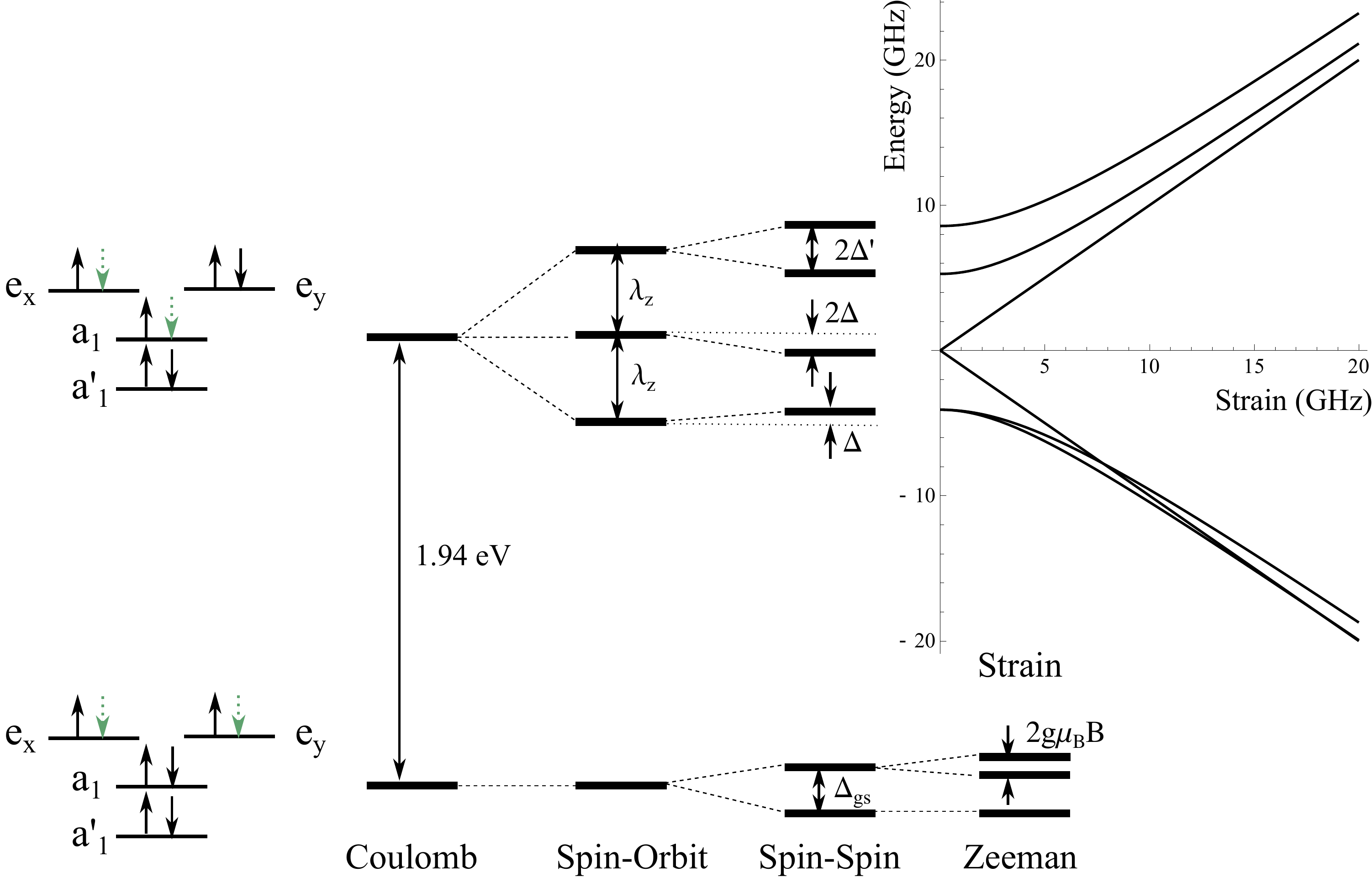}
\caption{Ground and excited states associated with the NV center's optical transitions. The four symmetrized states are filled by six electrons, which can also be viewed as two holes (indicated with green arrows). The Coulomb interaction defines the optical transition energy of the NV center. The effect of the spin-orbit and spin-spin interactions are indicated for both the ground and excited states. In the ground state, the application of a magnetic field further splits the $\ket{m_s=\pm1}$ states due to the Zeeman interaction. In the excited state, crystal strain modifies the energies of sublevels. See (Maze et al., 2011) for details.}
\label{f:electronicStates}
\end{figure}

In the ground state of the NV center, the $a'_1$ and $a_1$ states, which are lowest in energy,
are filled by four electrons, as shown in Figure \ref{f:electronicStates}. The remaining two
electrons occupy the degenerate orbitals $e_x$ and $e_y$.
The orbitals $e_x$ and
$e_y$ can be viewed as $p$-type orbitals and $e_+ = -e_x-ie_y$, $e_- =
e_x - ie_y$ are analogous to $p$ states with definite orbital angular
momentum.
An antisymmetric combination of the orbital states minimizes the Coulomb energy and results in a spin-triplet ground state manifold
\bea
\ket{^3A_2} = \ket{E_0}\otimes
\left\{ 
\begin{array}{c}
\ket{+1} \\ \ket{0}\\ \ket{-1}\end{array}
\right.
\eea
where $\ket{\pm1}, \ket{0}$ correspond to the $m_s=\pm1, 0$ states,
respectively. The notation $\ket{E_0} = \ket{e_x e_y-e_y e_x}$ indicates that the orbital state has 0 orbital angular momentum projection 
along the NV
axis.
$A_2$ denotes the orbital symmetry of the state, which
is determined by the symmetries of the $e_x$ and $e_y$ orbitals. From here on, we will often denote these ground states simply as $\ket{0}$ and $\ket{\pm1}$ when it's clear that we are referring to the full electronic state and not just the spin projections. 

The relevant excited state for the optical transitions of the NV center is a pair of triplets which arises from the
promotion of one of the electrons occupying the orbital $a_1$ to the
$e_x$ or $e_y$ orbitals \shortcite{Goss:PRL1996}. Note that this state can
be modeled by one hole in the orbital $e$ and another hole in the
orbital $a_1$, i.e. a triplet in the $ae$ electronic configuration
(similarly, the ground state can be modeled by two holes in the $e^2$
electronic configuration). A total of six states can be formed in this
configuration and their symmetries are determined by a group
theoretical analysis \shortcite{MazeNJP2011},
\begin{eqnarray}\label{eq:ae}
\ket{A_1} &=& \ket{E_-}\otimes\ket{+1}-\ket{E_+}\otimes\ket{-1} \nonumber \\
\ket{A_2} &=&  \ket{E_-}\otimes\ket{+1}+\ket{E_+}\otimes\ket{-1} \nonumber  \\
\ket{E_x} &=& \ket{X}\otimes\ket{0}  \nonumber \\
\ket{E_y} &=& \ket{Y}\otimes\ket{0}   \\
\ket{E_1} &=&  \ket{E_-}\otimes\ket{-1} - \ket{E_+}\otimes\ket{+1} \nonumber  \\
\ket{E_2} &=&  \ket{E_-}\otimes\ket{-1} + \ket{E_+}\otimes\ket{+1} \nonumber
\end{eqnarray}
where we have named the first four states as $A_1$, $A_2$, $E_x$ and
$E_y$ according to their symmetries and named the last two states as
$E_1$ and $E_2$ since they also transform according to the irreducible
representation $E$. Here, $\ket{E_\pm} = \ket{ae_{\pm} - e_{\pm}a}$
and $\ket{X(Y)} = \ket{ae_{x(y)}-e_{x(y)}a}$. 

As spin-orbit and spin-spin interactions are invariant under any
operation of the $C_{3v}$ group, the states given in equation
(\ref{eq:ae}) are eigenstates of the full Hamiltonian including these
interactions and in the absence of any perturbation such as magnetic field
and/or crystal strain. The Hamiltonians for the spin-orbit and spin-spin interactions can be written in the basis of the six states listed above as
\begin{eqnarray}
H_{so} &=& \lambda_z \left(\ketbra{A_1}{A_1}+\ketbra{A_2}{A_2}-\ketbra{E_1}{E_1}-\ketbra{E_2}{E_2} \right)\\ 
H_{ss} & =
& \Delta\rp{\ketbra{A_1}{A_1}+\ketbra{A_2}{A_2}+\ketbra{E_1}{E_1}+\ketbra{E_2}{E_2}}\nonumber \\
&&-2\Delta\rp{\ketbra{E_x}{E_x}+\ketbra{E_y}{E_y}}+\Delta^\prime\rp{\ketbra{A_2}{A_2}-\ketbra{A_1}{A_1}}\nonumber \\
&&+\Delta^{\prime\prime}\rp{\ketbra{E_1}{E_y}+\ketbra{E_y}{E_1}-i\ketbra{E_2}{E_x}+i\ketbra{E_x}{E_2}}
\end{eqnarray}
where $\lambda_z $
is the axial spin-orbit interaction and $3\Delta\approx1.42$ GHz and $\Delta^\prime \approx 1.55$ GHz
characterize the spin-spin induced zero-field splittings \shortcite{Lenef:PRB1996,Rogers:NJP2009,MazeNJP2011}. 
It can be seen
that spin-orbit interaction splits states with different total angular momentum,
i.e., the pairs $\rp{A_1,A_2}$, $\rp{E_x,E_y}$ and $\rp{E_1,E_2}$ are
split from each other by about 5.5
GHz \shortcite{Lenef:PRB1996,Manson2006,BatalovPRL2009}. The $\Delta^{\prime\prime}$ term leads to non-spin preserving cross transitions, which results to optical pumping of the spin states and plays a role in, for example, limiting the efficiency of spin readout using resonant excitation. The spin-spin interaction also
plays a crucial role in the stability of the states $\ket{A_1}$ and
$\ket{A_2}$. This interaction shifts up the non-zero spin states
($A_1$, $A_2$, $E_1$ and $E_2$) by $\sim\frac{1}{3}1.42$
GHz \shortcite{Fuchs:PRL2008,BatalovPRL2009} and shift the states
$\ket{E_x}$ and $\ket{E_y}$ down by $\sim\frac{2}{3}1.42$ GHz. Thus,
the gap between $\rp{A_1,A_2}$ and $\rp{E_x,E_y}$ is increased, but
the gap between the states $\rp{E_x,E_y}$ and $\rp{E_1,E_2}$ is
reduced. In addition, the spin-spin interaction splits the states
$\ket{A_2}$ and $\ket{A_1}$ by $\sim3.3$
GHz \shortcite{BatalovPRL2009}. Thus, for relatively low strain, we obtain
a very robust $\ket{A_2}$ state with stable symmetry properties that
are protected by an energy gap arising from the spin-orbit and
spin-spin interactions. 

We note here that the zero-field splitting in the ground state also arises from the spin-spin interaction, and gives an energy difference of $\Delta_{gs}\sim$2.88 GHz between the $\ket{m_s=0}$ and $\ket{m_s=\pm1}$ states.

\subsection{Properties of optical transitions}
Once the wavefunctions are known, it is possible to calculate the
selection rules of optical transitions between the triplet excited
state and the triplet ground state. The dipole moment between the
ground and excited state is produced by the hole left in the $a$
orbital under optical excitation. As mentioned earlier, the matrix elements $\bra{a}\hat{x}\cdot
\vec{r}\ket{e_x}$ and $\bra{a}\hat{y}\cdot \vec{r}\ket{e_y}$ are non-zero, where
$\hat{x}$ and $\hat{y}$ represent the polarization of the involved
photon. We can then calculate the selection rules for transitions
between every pair of ground and excited states, as shown in
Table \ref{t:rules}. As expected, these selection rules conserve the total angular momentum of the photon-NV center system.

\begin{table}
\caption{Selection rules for optical transitions between the triplet excited state $(ae)$ and the triplet ground state $(e^2)$. Linear polarizations are represented by $\hat{x}$ and $\hat{y}$, while circular polarization are represented by $\hat{\sigma}_\pm = \hat{x}\pm i\hat{y}$. As an example, a photon with $\sigma_+$ polarization is emitted when the electron decays from state $A_2$ to state $^3A_{2-}$. }
\begin{center}
\begin{tabular}{c|cccccc}
Pol & $A_{1}$ & $A_{2}$ & $E_{1}$ & $E_{2}$ & $E_{x}$ &
$E_{y}$ \\ \hline $^3A_{2-}$ & $\hat{\sigma}_+$ & $\hat{\sigma}_+$ &
$\hat{\sigma}_-$ & $\hat{\sigma}_-$ & & \\ $^3A_{20}$ & & & & &
$\hat{y}$ & $\hat{x}$ \\ $^3A_{2+}$ & $\hat{\sigma}_-$ &
$\hat{\sigma}_-$ & $\hat{\sigma}_+$ & $\hat{\sigma}_+$ & &
\end{tabular}
\end{center}
\label{t:rules}
\end{table}%

The allowed optical transitions and their polarization properties
indicate several possible $\Lambda$ schemes in the NV center. However,
while these properties are relatively robust in the $\ket{A_1}$ and
$\ket{A_2}$ states, the $\ket{A_1}$ state is coupled non-radiatively
to a metastable singlet state, which then decays to the
ground state $\ket{0}$. This results in leakage out of the $\Lambda$ system consisting of the $\ket{A_1}$ state and the $\ket{\pm1}$ states. Thus, we find that the $\ket{A_2}$
state provides the an almost ideal closed $\Lambda$ scheme for experiments such as spin-photon entanglement, as described in Section \ref{EntanglementChapter} and \shortcite{ToganNature2010}. On the other hand, the open $\Lambda$ system involving $\ket{A_1}$ can be useful for efficient detection of dark states during coherent population trapping, as shown in Section \ref{CPTChapter} and \shortcite{ToganNature2011}.

\subsection{The effect of strain}
\begin{figure}
\centering
\includegraphics[scale = 0.675]{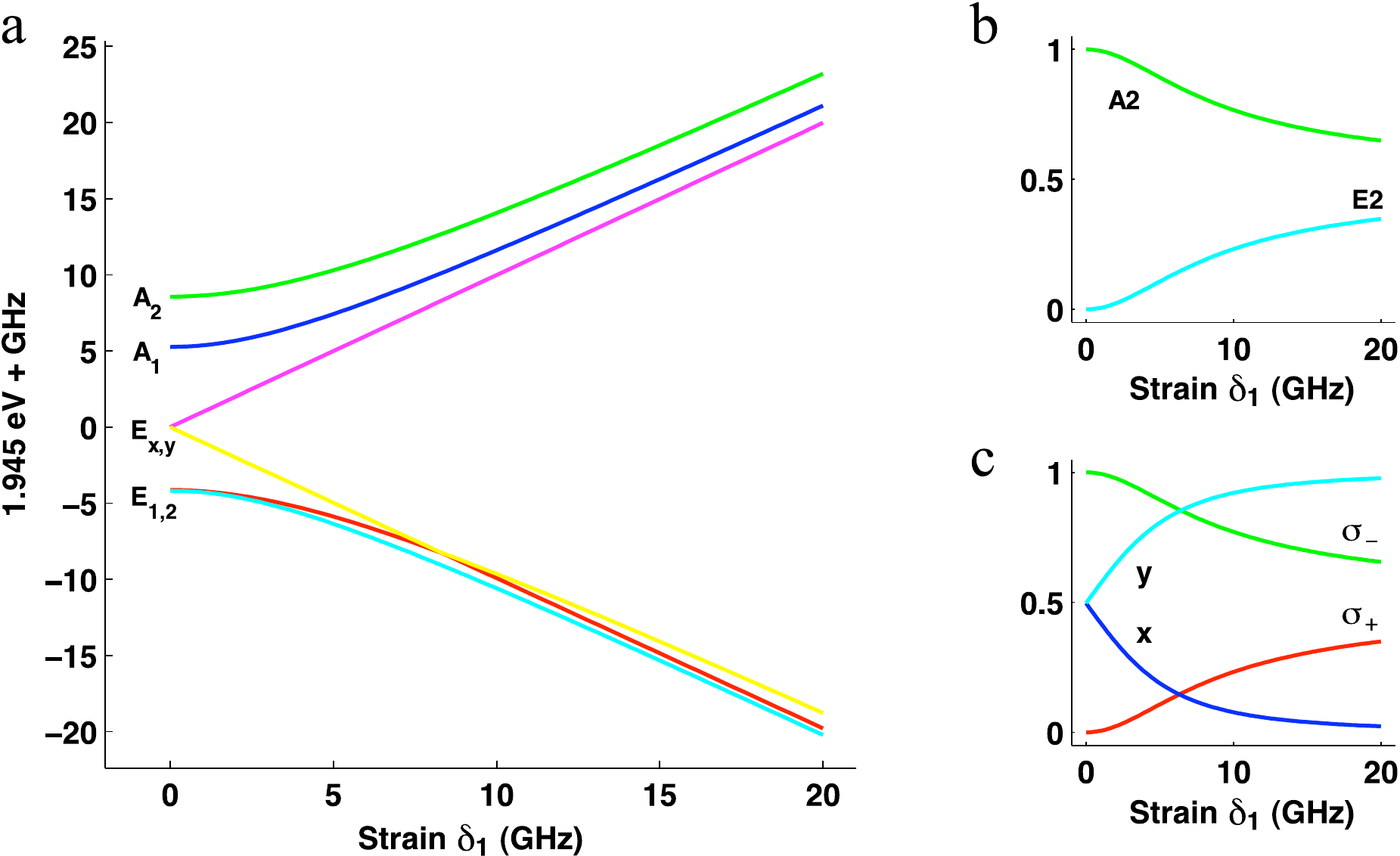}
\caption{Effect of strain on the properties of the NV center. \textbf{a.} Energies of the excited states as a function of strain, expressed in units of the linear strain induced splitting between the $\ket{E_x}$ and $\ket{E_y}$ states. \textbf{b.} Strain induced mixing of the $\ket{A_2}$ and $\ket{E_2}$ states, showing the fraction of $A_2$ and $E_2$ character of the highest energy excited state. \textbf{c.} Polarization character of the optical transition between $\ket{A_2}$ state and the ground state $\ket{+1}$ as a function of strain. As strain increases, the polarization changes from circularly to linearly polarized. The transition between $\ket{A_2}$ and $\ket{-1}$ shows the same behavior, except with $\sigma_-$ and $\sigma_+$ switched. Therefore, at high strain, decay from the $\ket{A_2}$ results in a separable state of the photon polarization and spin rather than an entangled state.}
\label{f:strain}
\end{figure}

We now discuss the effect of local strain on the
properties of the optical transitions in order to understand variations between different NV centers and deviations from the unperturbed system described above. This perturbation splits the
degeneracy between the $e_x$ and $e_y$ orbitals and results in their
mixing. In the limit of high strain (larger than the spin-orbit
splitting), the excited state manifold splits into two triplets, each
with a particular well defined spatial wavefunction. Orbital and spin
degrees of freedom separate in this regime and spin-preserving
transitions are excited by linearly polarized light.
The strain Hamiltonian is given by \shortcite{Rogers:NJP2009,MazeNJP2011}
\begin{eqnarray}
H_{strain} &=& \delta_1\rp{\ketbra{e_x}{e_x} - \ketbra{e_y}{e_y}} + \delta_2\rp{\ketbra{e_x}{e_y}+\ketbra{e_y}{e_x}}
\end{eqnarray}
where $\delta_1$ and $\delta_2$ are different parameters describing the crystal strain. Figure \ref{f:strain} shows an example of
how $\delta_1$ strain splits and mixes the excited states, and as a
results changes the polarization properties of the optical
transitions.  A similar effect occurs for $\delta_2$ strain, where
$\ket{A_2}$ is mixed with $\ket{E_1}$.

The full Hamiltonian for the excited state manifold of the NV center is then
\begin{eqnarray}
H &=& H_{ss} + H_{so} + H_{strain},
\end{eqnarray}
which can be written in the set of basis states $\left\{\ket{A_1}, \ket{A_2}, \ket{E_x}, \ket{E_y}, \ket{E_1}, \ket{E_2} \right\}$ as
\begin{equation}
H=\left(
\begin{array}{cccccc}
\Delta-\Delta'+\lambda_z & 0 & 0 & 0 & \delta_1 & -i\delta_2 \\
0 & \Delta+\Delta'+\lambda_z & 0 & 0 & i \delta_2 & -\delta_1 \\
0 & 0 & -2\Delta+\delta_1 & \delta_2 & 0 & i\Delta^{\prime\prime} \\
0 & 0 & \delta_2 & -2\Delta-\delta_1 & \Delta^{\prime\prime} & 0 \\
\delta_1 & -i\delta_2 & 0 & \Delta^{\prime\prime} & \Delta-\lambda_z & 0 \\
i \delta_2 & -\delta_1 & -i\Delta^{\prime\prime} & 0  & 0 & \Delta-\lambda_z
\end{array}
\right)
\end{equation}
\label{e:Hamiltonian}

It is worth mentioning that while the excited state configuration is
highly affected by strain, the ground state configuration is
unaffected to first order due to its antisymmetric combination of
$e_x$ and $e_y$ orbitals. It is also protected by the large optical
gap between the ground and excited state to second order perturbation
in strain. 

\section{Experimental techniques}

Having developed a theoretical model for the optical transitions of the NV center, we now move on to experimental explorations of these properties. While there are many available descriptions of setups and procedures for room temperature experiments with NV centers \shortcite{Gruber97,ChildressScience2006}, we will focus here on some techniques relevant for low-temperature studies of the optical properties. 

\subsection{Diamond materials}

We begin this section with a description of different types of single crystal diamond samples and their properties with regard to NV center properties. The first type of commonly available diamond material is classified as type Ib. In these diamonds, substitutional nitrogen atoms (also called P1 centers) are the dominant defects, which cause the samples to have a yellow color \shortcite{Zaitsev10}. Type Ib diamonds can occur naturally and are typically grown by the high-pressure-high-temperature (HPHT) method by companies such as Element Six and Sumitomo. Element Six specifies their type Ib diamond as having substitutional nitrogen concentrations of $<$200 ppm. In our experience, the concentration of NV centers in type Ib samples vary widely through different samples, but NV centers in bulk samples are usually unresolvable under confocal microscopy. There have been no reports of coherent optical transitions from type Ib diamonds, except in one nanocrystal sample where a narrow, but spectrally unstable line was observed under photoluminescence excitation (PLE) spectroscopy \shortcite{ShenPRB2008}. However, as far as we know, this result has never been reproduced. In our experiments, type Ib samples are typically used for developing and fabricating nanophotonic structures and studies of NV centers in these structures at room temperature.

\begin{figure}
\centering
\includegraphics[width=\textwidth]{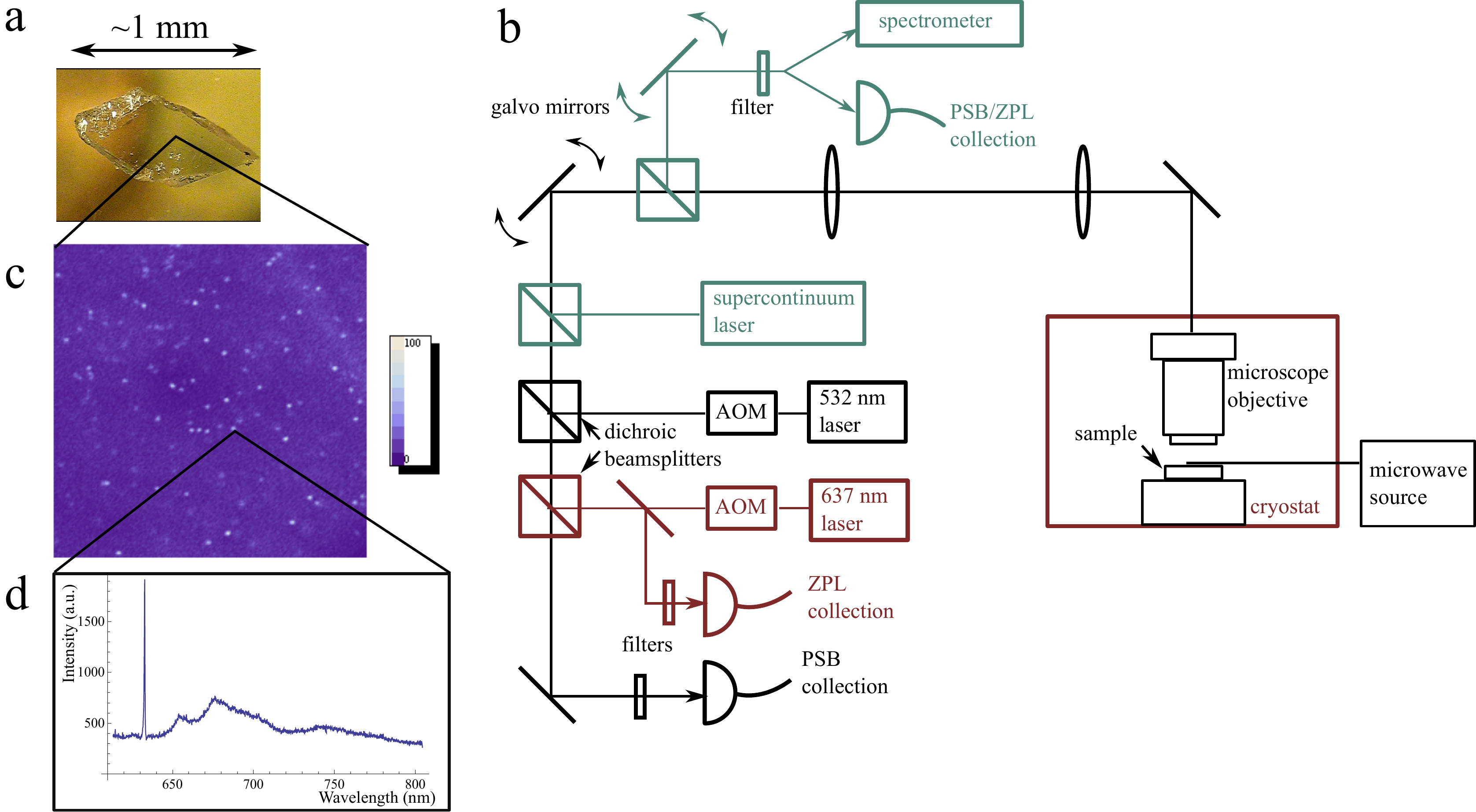}
\caption{Experimental setup and characterization of NV centers. \textbf{a.} The type IIa natural diamond sample used for the experiments in Sections \ref{EntanglementChapter} and \ref{CPTChapter}. \textbf{b.} Confocal microscope setup. Components in black are needed for a basic room temperature apparatus for investigating the ground state spin physics of the NV center. Red components, such as a cryostat and resonant excitation laser, are additionally needed for low temperature studies of coherent optical properties. A supercontinuum laser is added for characterization of photonic devices such as cavities and waveguides, and an additional, independent collection channel is used for collecting light from a different spatial location than the excitation, as shown in green. In addition, the spectrometer can also be used for photoluminescence studies for the NV center. \textbf{c.} Confocal image of single NV centers obtained using 532 nm excitation and PSB collection. \textbf{d.} Low temperature photoluminescence spectrum of a NV center. Sharp peak is ZPL at 637 nm.}
\label{f:setup}
\end{figure}

The rest of the diamond samples that are suitable for the experiments described here can be classified as type IIa, which is typically defined as having nitrogen impurity concentrations of less than 1ppm. Type IIa diamonds are rare in nature, but one such sample was found containing a resolvable concentration of single NV centers that show optical linewidths that are close to the lifetime limited linewidth of 13 MHz \shortcite{TamaratPRL2006}. Many initial experiments on the optical properties of NV centers were performed on several different pieces of this sample, including the ones described in Sections \ref{EntanglementChapter} and \ref{CPTChapter} of this chapter, along with, for example, \shortcite{TamaratPRL2006} and \shortcite{BatalovPRL2008}. The unique properties of this sample, and its somewhat exotic origins in the Ural mountains, has earned it the nickname ``The Magic Russian Diamond". The piece used in our experiments is shown in Figure \ref{f:setup}a. Clearly, such specialized samples cannot be used for wide-scale experiments and development of diamond-based devices. An alternative is to work with synthetic type IIa diamonds grown using microwave assisted chemical vapor deposition (CVD) at companies such as Element Six Inc. Standard type IIa grade samples typically also show a unresolvable concentration of NV centers. However, ``electronic grade" diamonds developed by Element Six, which are specified to have substitutional nitrogen concentrations of $<$5 ppb, have NV center concentrations that range from barely resolvable single NVs to less than 1 NV per tens of $\mu$m$^3$. Native NV centers found in these samples often have linewidths that are broadened by spectral diffusion to several hundred MHz \shortcite{BernienPRL2012,AcostaPRL2012,StaceyAM2012}. However, they have been used in a variety of experiments involving optical transitions of NV centers \shortcite{BernienNature2013,BernienPRL2012,Sipahigil2012}, and are also excellent starting materials for introducing optically coherent NV centers through ion implantation, as described in Section \ref{ImplantationChapter}.

\subsection{Experimental setup for optical characterization of NV centers}

Figure \ref{f:setup}b shows a schematic of a multi-purpose setup used for optical investigation of NV centers and diamond-based photonic devices. Depending on the type of experiment, only parts of this comprehensive setup are needed. A typical basic confocal microscope for room-temperature studies of NV centers is shown in black. A high numerical aperture (NA) objective is used to focus laser light onto the sample. In order to scan over the sample and locate individual centers in a confocal image such as shown in Figure \ref{f:setup}c, the diamond is either mounted on a high resolution positioning stage, or the optical path is scanned using a galvo mirror imaged onto the back of the objective with a pair of lenses comprising a 4f imaging system. The vibronic sidebands of the NV center's optically excited states can be addressed using a laser of higher energy than the 637 nm zero-phonon line (ZPL). Typically, this is done using a 532 nm frequency doubled Nd:YAG laser, which is the most common method for controlling the charge state of the NV center. CW 532 nm excitation can ionize both the negatively charged NV center and surrounding charge donors such as substitutional nitrogens, resulting in a equilibrium charge state of the NV center that is $\sim70\%$ NV$^-$ and $\sim30\%$ NV$^0$ \shortcite{Aslam2013}. Off resonant excitation is also used in room temperature experiments to polarize the spin of the NV center in to the $\ket{m_s=0}$ ground state. This process happens through non-spin preserving transitions from the optically excited state into spin-singlet states of the NV center, which then preferentially decays to the $\ket{m_s=0}$ state \shortcite{Manson2006}. Finally, microwave and RF fields for driving spin transitions of the NV centers can be applied with a variety of methods, including copper wires, coplanar waveguides, and resonator structures. 

For low temperature experiments, the diamond sample is placed inside a helium flow cryostat (Janis ST-500) and cooled to $\sim$4K. Alternative low temperature systems such as closed-cycle pulse tube coolers and bath cryostats can also used. Diamond samples are typically mounted onto the cold finger, or some intermediate substrate such as a silicon wafer, using indium or silver paste. For resonant excitation of the NV centers, a external cavity diode laser around 637 nm is introduced into the optical path. The New Focus TLB 6304 laser is a popular choice due to its large mode-hop-free scanning range of 60-80 GHz, wavelength tunability over several nanometers, and ease of operation. Both the 532 nm and 637 nm lasers need to be pulsed for most experiments, which can be accomplished using acoustic-optical modulators (AOM) or electro-optical modulators (EOM).

Fluorescence from the NV center is collected through the same optical path, coupled into single-mode fibers, and detected using avalanche photodiodes (APD) or a spectrometer. Filters are placed in the collection path to remove reflected excitation light and separate different parts of the emission spectrum. As shown in Figure \ref{f:setup}d, the NV center spectrum consists of a broad separate phonon-sideband (PSB) and a sharp ZPL at low temperatures. The percentage of emission into the ZPL is typically $3-5\%$, which is a major limitation on the efficiency of applications based on coherent photons from the NV center. In Section \ref{NanophotonicsChapter}, we will address this issue by enhancing the emission into the ZPL using cavity QED with nanophotonic structures. In order to characterize these structures, we use an additional supercontinuum laser as a broadband light source to perform transmission measurements of cavities and waveguides. The light is coupled into one port of the optical device, and the transmitted light must be collected from an output port at a spatially separate location. For this purpose, a second collection channel can be added with its own set of galvo mirrors. In addition, the signal from the two channels can be combined on a time-correlated single photon counting (TCSPC) device for measurements of the second-order autocorrelation function $g^{2}(\tau)$ of the emission from individual NV centers to show that they are single-photon emitters \shortcite{Kurtsiefer2000}. This can also be accomplished by splitting the light collected in a single channel using a fiber beamsplitter.

\section{PLE spectroscopy of NV centers}

\begin{figure}
\centering
\includegraphics[width=\textwidth]{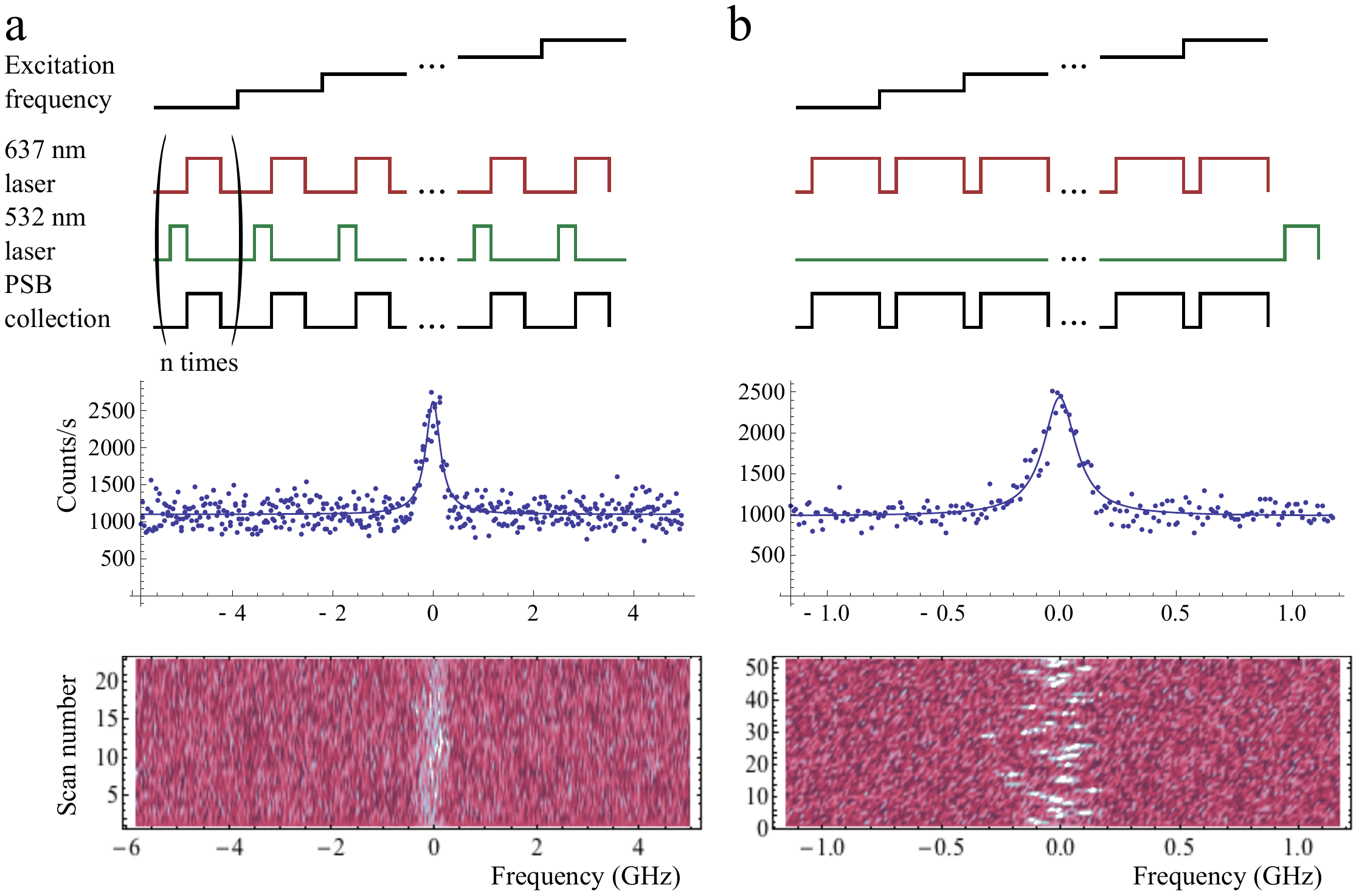}
\caption{Methods for PLE spectroscopy of NV centers. \textbf{a.} Pulse sequence for repumping during each PLE scan (top), along with an example showing the result of successive scans (bottom) and the averaged spectrum (middle). \textbf{b.} Pulse sequence for repumping at the end of each PLE scan (top), along with corresponding results for the same transition as in a., except zoomed in over a smaller frequency range.}
\label{f:PLE}
\end{figure}

The optical transitions of the NV center can be identified using PLE spectroscopy by scanning the resonant excitation laser across the ZPL and collecting photons in the PSB. Since the $\ket{0} \rightarrow \ket{E_x}$ and  $\ket{0} \rightarrow \ket{E_y}$ are best the cycling transitions, they are usually the ones observed in PLE spectroscopy in the absence of microwave driving or laser modulation. However, repeated resonant excitation eventually ionizes the NV$^-$ and optically pumps the NV center into a different ground state spin sublevel due to the small non-spin conserving cross transitions out of the $\ket{m_s=0}$ excited states into the $\ket{\pm1}$ ground states \shortcite{Tamarat:NJP2008}. Therefore, some form of charge and spin repumping must eventually be performed. As shown in Figure \ref{f:PLE}b, the first method we use involves scanning the 637 nm laser across the NV transitions and turning on the laser while simultaneous collecting PSB counts for $\sim2$ ms at each frequency. At the end of the scan, the 637 nm laser is turned off and a 532 nm repumping pulse is applied. Any duration of repumping longer than a few $\mu$s is sufficient to reinitialize the NV center. Depending on the charge environment of the NV center and its strain-dependent level structure, however, it may be difficult to obtain good signal from this simple method. For example, in the presence of a large number of other charge traps in the environment, the NV center may be ionized before a sufficient number of photons are scattered on resonance. Another potential issue is that the $\ket{0} \rightarrow \ket{E_x}$ and  $\ket{0} \rightarrow \ket{E_y}$ transitions have some branching ratio into the $m_s=\pm1$ states, causing a decrease in fluorescence on the timescale of several microseconds. However, fluorescence can potentially still be detected due to off-resonant excitation of transitions from the $\ket{\pm1}$ spin states. For example, at high strain, the $\ket{\pm1}\rightarrow\ket{A_1}$ transition becomes almost degenerate with the $\ket{\pm1}\rightarrow\ket{E_x}$ transition, allowing a steady state fluorescence to be established. However, this may not be the case for all NV centers. More robust methods of obtaining the PLE spectrum without repumping during the scan involve either CW microwave mixing of the $\ket{0}$ and $\ket{\pm1}$, or modulation of the resonant excitation laser by the ground state zero-field splitting, which also allows the $\ket{m_s=\pm1}$ transitions to be observed \shortcite{Tamarat:NJP2008,FuPRL2009,SantoriPRL2006}. 

Figure \ref{f:PLE}b shows the results of PLE scans taken using this first method. A clear resonance can be seen in each scan, whose frequency then jumps during each successive scan. This is an indication of spectral diffusion, where changes in the local electric field environment of the NV center causes the energies of its excited states to fluctuate over time. As mentioned earlier, from the electronic orbitals $\ket{e_x}$ and $\ket{e_y}$ involved in the excited states, one can see that the $\ket{E_x}$ and $\ket{E_y}$ states have a permanent electric dipole moment. This dipole moment has been measured to be $\sim$0.8 Debye, which gives rise to energy shifts of $\sim$4 GHz/MV/m \shortcite{MazeNJP2011}. It is now well established that 532 nm repumping causes complex charge dynamics in the diamond environment by photoionizing impurities that act as charge donors \shortcite{BassettPRL2011}. Therefore, while very narrow single-scan linewidths can be obtained using this method \shortcite{ShenPRB2008,TamaratPRL2006,FuPRL2009}, the long-term, extrinsically broadened linewidth obtained by averaging many such scans can be much broader \shortcite{BernienPRL2012,AcostaPRL2012,StaceyAM2012}. This issue is addressed in more detail in Section \ref{ImplantationChapter} of this chapter.
 
A second method of obtaining the PLE spectrum involves a pulse sequence consisting of a 1 $\mu$s 532 nm repumping pulse, followed by a 10 $\mu$s resonant excitation pulse, during which PSB counts are collected. As the 637 nm laser is scanned across the NV transitions, this sequence is repeated many times for each frequency, as illustrated in Figure \ref{f:PLE}. This method has the advantage that it is robust against ionization and spin polarization, and should in principle give good signal to noise regardless of the charge environment and strain-dependent level structure of the NV center. It also shows the overall lineshape of the transition, including effects of spectral diffusion, in a single scan. Again, microwave fields are used to obtain a spectrum of all the possible transitions, including those that involve the $\ket{m_s=\pm1}$ states. 

\section{Optical properties of NV centers}

\begin{figure}
\centering
\includegraphics[width=\textwidth]{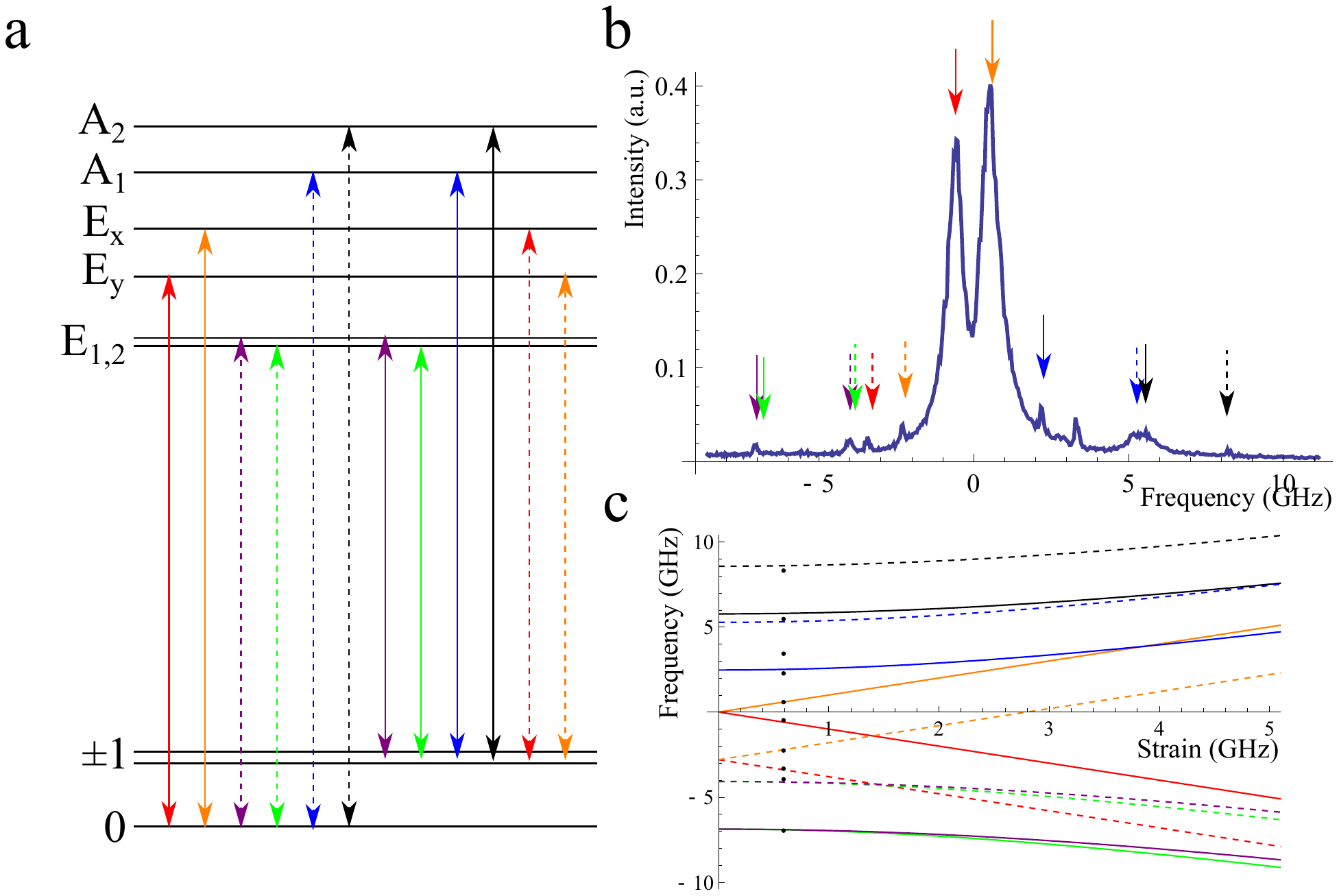}
\caption{Optical transitions of the NV center. \textbf{a.} All possible transitions between the ground and excited states, with direct transitions indicated with solid lines, and spin non-conserving cross transitions indicated with dashed lines. \textbf{b.} PLE spectrum taken with CW microwave excitation. \textbf{c.} Frequencies of all possible transitions shown in a. as a function of strain. The frequencies of the peaks in b. are matched to a particular strain value. The extra unidentified peak may be due to a two photon transition from the $\ket{0}$ state to the $\ket{E_x}$ state through the absorption of an optical photon and emission of a microwave photon.}
\label{f:allTransitions}
\end{figure}

Figure \ref{f:allTransitions}a shows a PLE spectrum taken with CW microwave excitation that mixes the ground states during the 637 nm laser pulse. All resonances expected from the direct and cross transitions are observed, as shown in \ref{f:allTransitions}b, along with some possible microwave-optical two photon transitions.

%\section{Optical Rabi oscillations and resonant spin readout}

\begin{figure}
\centering
\includegraphics[width=\textwidth]{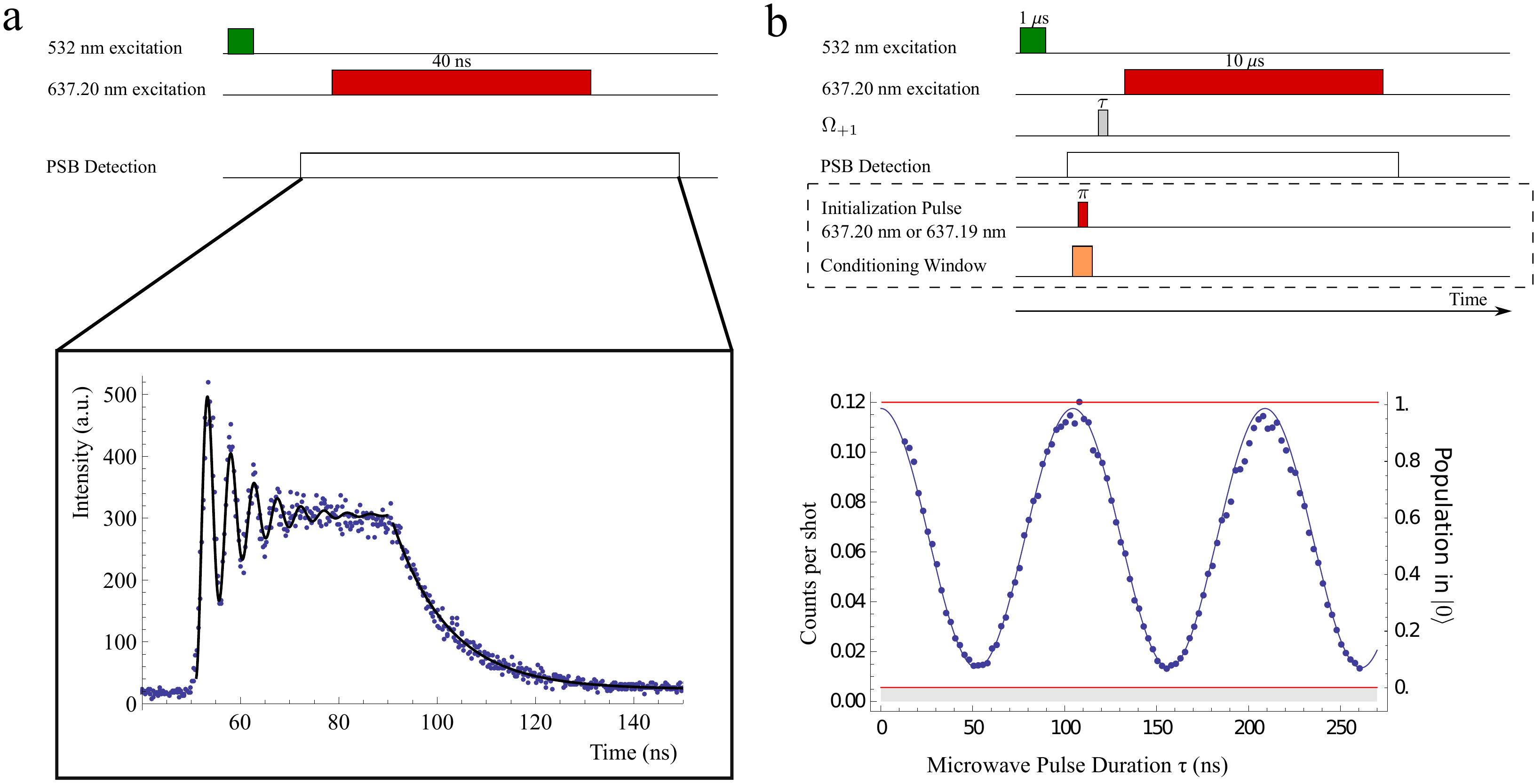}
\caption{\textbf{a.} Optical Rabi oscillations detected using PSB fluorescence and resonant excitation, along with corresponding pulse sequence. The 637.2 nm laser is tuned to the $\ket{0} \rightarrow \ket{E_y}$ transition and is on from 50 ns to 90 ns. \textbf{b.} Rabi oscillations between $\ket{0}$ and $\ket{+1}$ (blue) detected using resonant spin readout and corresponding pulse sequence. Red lines are calibration levels for the $\ket{0}$ and $\ket{+1}$ states. To initialize the spin states for this calibration, we use a measurement-based technique where the $\ket{0}$ is prepared conditioned on the detection of a PSB photon after excitation to the $\ket{E_y}$ state. To prepare the $\ket{+1}$ state, we then apply a microwave $\pi$ pulse before readout.}
\label{f:rabiReadout}
\end{figure}

The strongest transitions in \ref{f:allTransitions}a are the $\ket{0} \rightarrow \ket{E_x}$ and  $\ket{0} \rightarrow \ket{E_y}$ cycling transitions. Figure \ref{f:rabiReadout}a shows an example of optical Rabi oscillations on the $\ket{0} \rightarrow \ket{E_y}$ transition, measured by turning on a 40 ns resonant laser pulse a rise time of $\sim$1 ns using an EOM (Guided Color Technologies). We then accumulate a histogram of photon arrival times on a TCSPC device with a timing resolution of 195 ps. The photon count rate is proportional to  the population in the $\ket{E_y}$ state, which, in the absence of decay or decoherence, is given by $P_{E_y}(t)=$cos$^2(|\Omega|t)$, where $\Omega=\vec{\mu}\cdot\vec{E}/\hbar$ is the Rabi frequency. In this case, however, the oscillations are damped by the finite lifetime of the excited states, which can be measured by fitting to the exponentially decaying tail after the pulse turns off. In addition to population decay, additional decoherence, such as frequency fluctuations of the transition frequency relative to the laser frequency caused by, for example, spectral diffusion, will also contribute to the damping of Rabi oscillations. We note here, however, that this additional dephasing alone does not always give a good estimate of the spectral diffusion of the NV center, unlike what is sometimes claimed in literature. For example, if the transition frequency jumps well outside of the power-broadened linewidth of the transition, then the NV center will effectively not be excited, and there will simply be no contribution to the Rabi oscillation data. 

By measuring Rabi oscillations, we can determine the length of laser pulse needed to, for example, perform an optical $\pi$ pulse to transfer population to the $\ket{A_2}$ state for entanglement generation, as described in Section \ref{EntanglementChapter}. Note that, while optical Rabi oscillations can also be observed between the $\ket{m_s=\pm1}$ ground and excited states, the dynamics are more complicated due to the fact that these are $\Lambda$-type transitions. In the absence of a magnetic field, instead of settling to a steady state value corresponding to 50$\%$ population in the excited state, the photon count rate continually decreases after a few optical cycles as population is pumped into a "dark" ground state that is not excited by the by the laser. However, the observed Rabi oscillations are usually sufficient for determining the optical $\pi$ pulse length.
%These dynamics are explored in more detail in Chapter \ref{CPTChapter} .

The cycling transitions of the NV center can also be used to efficiently read out the spin state of the NV center. As shown in Figure \ref{f:rabiReadout}, when a resonant laser pulse is turned on for 10 $\mu$s, the number of photons scattered is proportional to the population in the $\ket{m_s=0}$ states, and microwave-driven Rabi oscillations between the $\ket{0}$ and $\ket{+1}$ ground state sublevels can be observed. A detailed description of the resonant spin-readout procedure used in Section \ref{EntanglementChapter} can be found in \shortcite{ToganNature2010}. The efficiency and fidelity of this spin readout method is limited by the photon collection efficiency, the temperature, and the branching ratio out of the $\ket{m_s=0}$ manifold. As can be predicted by the Hamiltonian given in Equation \ref{e:Hamiltonian}, this branching ratio depends on the local strain. In our first experiments, the spin readout contrast was limited to $c_M=0.11\pm0.0022$ counts/shot when the spin was prepared in the $\ket{0}$ state and a background-limited $c_B=0.0057\pm0.0010$ counts/shot when the spin was prepared in the $\ket{\pm1}$ states. Therefore, to determine the spin of the NV center, the state had to be prepared and read-out many times to obtain an average number of counts per shot $C$, which then allows us to calculate the population in the $\ket{0}$ state as $P=\left(C-c_B\right)/\left(c_M-c_B\right)$. By increasing the collection efficiency with a solid immersion lens and working with low strain NV centers, single shot resonant readout of the spin state has subsequently been demonstrated \shortcite{Lucio11}. 

The rest of the chapter gives several examples of applications based on the optical properties of NV centers that we have just described. 

\section{Quantum entanglement between an 
optical photon and a solid-state spin qubit}
\label{EntanglementChapter}

%The rest of the chapter gives several examples of applications based on the optical properties of NV centers. First, we describe the realization of quantum entanglement between a NV center and a single photon. The scheme we use here is inspired by previous work done in trapped ions \shortcite{Blinov:Nature2004}. Based on our knowledge of the NV center's level structure, as described in Section \ref{OpticalProperties}, we identify a $\Lambda$ type three-level structure that enables the generation of entanglement between the $m_s=\pm1$ spin sublevels of the NV center's ground state and the polarization degree of freedom of spontaneously emitted photons. In addition, we demonstrate resonant readout of the spin-state of the NV center using a different, cycling optical transition involving the $m_s=0$ spin states, which allows us to verify the fidelity of the entanglement. 

\subsection{Introduction}

A quantum network \shortcite{Kimble:Nature2008} consists of several nodes, each containing a
long-lived quantum memory and a small quantum
processor, that are connected via entanglement. Its potential applications include long-distance quantum communication and
distributed quantum computation \shortcite{Duan:AAMOP2008}.  Several recent experiments demonstrated on-chip
entanglement of solid-state qubits  separated by nanometer \shortcite{Neumann08} to millimeter \shortcite{Ansmann:Nature2009,DiCarlo:Nature2009} length scales. However, realization of long-distance entanglement
based on solid-state systems coupled to single optical photons \shortcite{Riedmatten:Nature2008} is an outstanding challenge.
%Connections between nodes can be achieved by coupling the
%memory qubit with the quantum light field.  
%Realization of quantum
%networks based on solid-state systems\shortcite{Imamoglu:PRL1999, Hanson:Nature2008} 
%is an outstanding challenge with
%potential applications for scalable quantum communication and
%computation.  
The NV
center is a promising candidate for
implementing a quantum node. The ground state of the negatively charged NV
center is an electronic spin triplet with a 2.88 GHz zero-field
splitting between the $\ket{m_s=0}$ and $\ket{m_s=\pm1}$ states (from
here on denoted $\ket{0}$ and $\ket{\pm1}$). 
%It can be manipulated and
%measured using optical and microwave radiation\shortcite{Fuchs:Science2009}. 
%The electron spin is a promising qubit candidate
%due to its long coherence times\shortcite{Balasubramanian:NMat2009}.  
With long coherence times \shortcite{Balasubramanian:NMat2009}, fast microwave manipulation, and optical preparation and detection \shortcite{Fuchs:Science2009}, the NV electronic spin presents a promising qubit candidate.
Moreover, it can be
coupled to nearby nuclear spins that provide exceptional quantum
memories and allows for the
robust implementation of few-qubit quantum
registers \shortcite{Neumann08,DuttScience2007}. In this section we
demonstrate %a key element of a quantum optical network,
%namely
the preparation of quantum entangled states between a single photon
and the electronic spin of a NV center:
\begin{equation}
|\Psi\rangle = \frac{1}{\sqrt{2}} (\ket{\sigma_-} \ket{+1} +
\ket{\sigma_+} \ket{-1}),
\label{equation1}
\end{equation} 
where $|\sigma_+\rangle$ and $|\sigma_-\rangle$ are orthogonal
circularly polarized single photon states.

\begin{figure}[htbp]
\begin{center}
\includegraphics[width = 0.5 \textwidth]{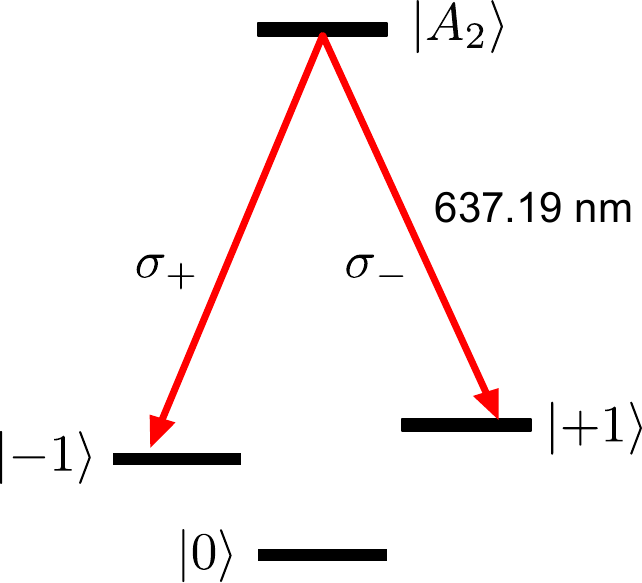}
\caption{Scheme for spin-photon entanglement. Following
  selective excitation to the $\ket{A_2}$ state, the $\Lambda$ system
  decays to two different spin states through the emission of
  orthogonally polarized photons, resulting in spin-photon
  entanglement. 
  }\label{EntangleFig1}
\end{center}
\end{figure}

The scheme we use here is inspired by previous work done in trapped ions \shortcite{Blinov:Nature2004}. The key idea of our experiment is illustrated in Figure~\ref{EntangleFig1}.
The NV center is prepared in a specific excited state ($\ket{A_2}$ in Fig~\ref{EntangleFig1}) that
decays with equal probability into two
different long lived spin states ($\ket{\pm 1}$) by the emission of orthogonally	
polarized optical photons at 637 nm. The entangled state given by Eq.  \ref{equation1} is created because photon polarization is uniquely correlated with the final spin state. This entanglement is verified by spin state measurement using a cycling
optical transition following the detection
of a 637 nm photon of chosen polarization.

\subsection{Experimental demonstration of spin-photon entanglement}

 We now turn to the experimental demonstration of spin-photon
 entanglement. The experimental setup is outlined in \shortcite{ToganNature2010}. To create the entangled state, we use coherent emission within the narrow-band zero phonon line (ZPL), which includes only 4\% of the NV center's total emission. The remaining optical radiation occurs in the frequency shifted phonon side band (PSB), which is accompanied by phonon emission that
 deteriorates the spin-photon entanglement \shortcite{KaiserArxiv2009}. Isolating the weak ZPL emission presents a
 significant experimental challenge due to strong reflections of the resonant
 excitation pulse reaching the detector.  
% We use a short (2 ns) resonant $\pi$ pulse for
% selective excitation of the $\ket{A_2}$ state. As the emission
% timescale is longer compared to the pulse length, the reflected photons and emission from the NV center may be
% separately identified in their detection time.  
By exciting the NV center with a circularly polarized 2 ns $\pi$ pulse that is shorter than the emission timescale, we can use detection timing to separate reflection from fluorescence photons.
A combination of confocal rejection,
 modulators, and finite transmittivity of our optics suppresses the
 reflections sufficiently to clearly detect the NV center's ZPL
 emission in a 20 ns region.
 
%
%\begin{figure}[htbp]
%\begin{center}
%\includegraphics[width = 0.9 \textwidth]{EntangleFigure3}
%\caption{Experimental procedure for entanglement
%  generation. \textbf{a.} After spin polarization into $\ket{0}$, population is transfered to $\ket{+1}$ by a
%  microwave $\pi$ pulse ($\Omega_{+1}$). The NV is excited to
%  $\ket{A_2}$ with a 637.19 nm $\pi$ pulse and the ZPL emission
%  is collected. \textbf{b.} If a $\sigma_+$ or $\sigma_-$ photon is
%  detected, the population in $\ket{+1}$ or $\ket{-1}$ is transferred
%  to $\ket{0}$. If a $\ket{H}$ or $\ket{V}$ photon is detected, a
%  $\tau-2\pi-\tau$ echo sequence (see Appendix~\ref{EntanglementAppendix}) is applied with $\Omega_{+1}$ and $\Omega_{-1}$, followed by a
%  $\pi$ pulse which transfers the population in $\ket{M}$ (see text) to
%  $\ket{0}$. \textbf{c.} The population in $\ket{0}$ is
%  measured using the 637.20 nm optical readout transition. \textbf{d.} Pulse sequence for the case
%  where a $\ket{H}$ or $\ket{V}$ ZPL photon is detected (time axis not to
%  scale). If a $\sigma_\pm$ photon is detected instead, only a $\pi$ pulse on either $\Omega_{+1}$ or $\Omega_{-1}$ is used for spin readout. Inset:
%  Detection time of ZPL channel photons showing reflection
%  from diamond surface and subsequent NV emission (blue) and background
%  counts (purple).}
%\label{EntangleFig3}
%\end{center}
%\end{figure}

\begin{figure}[htbp]
\begin{center}
\includegraphics[width = 0.9 \textwidth]{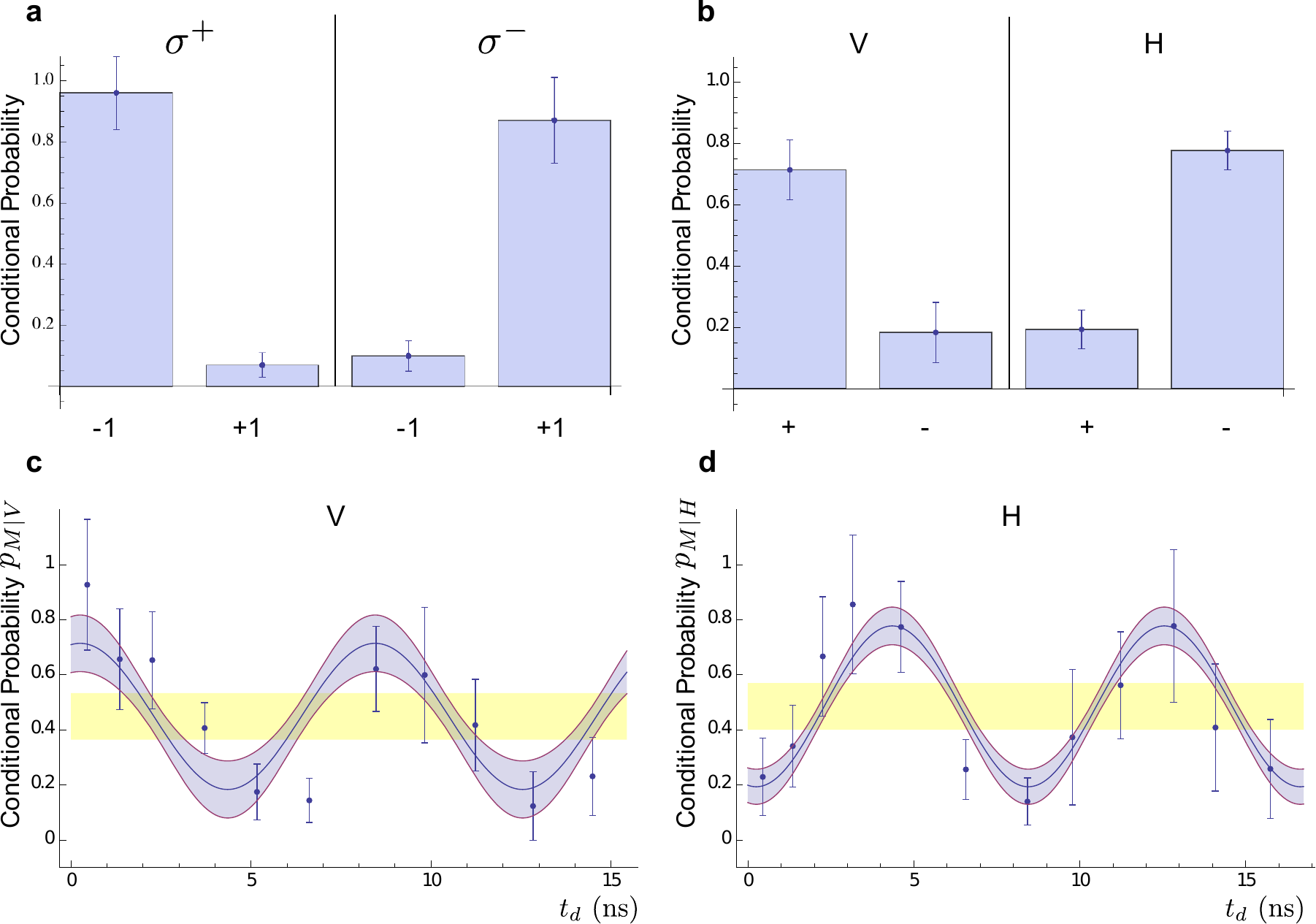}
\caption{Measurement of spin-photon correlations in two bases. \textbf{a.}
  Conditional probability of measuring
  $\ket{\pm1}$ after the detection of a $\sigma+$
  or $\sigma-$ photon.  \textbf{b.} Conditional probability of measuring
  $\ket{\pm}$ after the detection of a $H$ or $V$ photon, extracted from a fit to data shown in (c) and (d). \textbf{c\&d.} Measured conditional
  probability of finding the electronic spin in the state $\ket{M}$ after detection of a 
  $V$ (c) or $H$ (d) photon at time $t_d$.  Blue shaded region is the 68\% confidence interval for the fit (solid line) to the time-binned data. Errors on data points are one STD. Combined with the data shown in Figure~\ref{EntangleFig4}(a), oscillations with amplitude outside of the yellow regions result in fidelities greater than 0.5. The visibility of the measured oscillations are $0.59\pm0.18$ (c) and $0.60\pm0.11$ (d). Reprinted from (Togan \textit{et al.}, 2010).
%  \textbf{Inset:} Probability distribution of measured fidelity based on a maximum likelihood estimate described in  Appendix~\ref{EntanglementAppendix}, showing a 99.65\% probability of having demonstrated an entangled state.
%We detect on average one ZPL photon every few seconds. Since $\sim$100 repetitions are required for reliable spin state measurement, the data for each of the four photon polarizations shown were acquired over $\sim$24 hours.
}
\label{EntangleFig4}
\end{center}
\end{figure}

For photon state determination, 
ZPL photons in either the $\ket{\sigma_{\pm}}$ or $\ket{H} =
\frac{1}{\sqrt{2}}(\ket{\sigma_+} + \ket{\sigma_-})$, $\ket{V} =
\frac{1}{\sqrt{2}}(\ket{\sigma_+} - \ket{\sigma_-})$ basis are selected by a polarization analysis stage and detected after an optical path of $\sim2$ m. Spin
readout then occurs after a 0.5 $\mu$s spin memory interval following photon detection by transferring population from either
the $\ket{\pm1}$ states or from their appropriately chosen
superposition into the $\ket{0}$ state using microwave pulses
$\Omega_{\pm1}$. The pulses selectively address the $\ket{0}
\leftrightarrow \ket{\pm1}$ transitions with resonant frequencies
$\omega_{\pm}$ that differ by $\delta\omega=\omega_+-\omega_-=$122 MHz
due to an applied magnetic field. For superpositions of $\ket{\pm1}$
states an echo sequence is applied before the state transfer to extend
the spin coherence time. The transfer is followed by resonant
excitation of the $\ket{0} \leftrightarrow \ket{E_y}$ transition and collection of the PSB fluorescence. 
%To
%determine the measured population of the $\ket{0}$ state, we carefully
%calibrate fluorescence detected during the readout pulse using the
%procedure detailed in Methods.
We carefully calibrate the transferred population measured in the $\ket{0}$ state using the
procedure detailed in \shortcite{ToganNature2010}.
%We measure correlations between the appropriately polarized ZPL photons and their
%corresponding spin readout results. 
Figure~\ref{EntangleFig4}a
shows the populations in the $\ket{\pm1}$ states, measured
conditionally on the detection of a single circularly polarized ZPL photon.
Excellent
correlations between the photon polarization and NV
spin states are observed.

To complete the verification of entanglement, we now show that
correlations persist when ZPL photons are detected in a rotated polarization basis. 
Upon detection of a linearly polarized $\ket{H}$ or $\ket{V}$ photon at
time $t_{d}$, the entangled state in Eq.\ref{equation1} is
projected to $\ket{\pm}=\frac{1}{\sqrt{2}}\left(\ket{+1}\pm\ket{-1}\right)$, respectively. These states subsequently evolve in time ($t$) according to 
\begin{equation}
\ket{\pm}_t=\frac{1}{\sqrt{2}}\left(e^{-i\omega_+(t-t_d)}\ket{+1}\pm e^{-i\omega_-(t-t_d)}\ket{-1}\right).
\end{equation}
 In order to read out the relative phase of
superposition states between $\left\vert +1\right\rangle $ and
$\left\vert -1\right\rangle $, we use two resonant microwave fields with frequencies $\omega_{+}$ and $\omega_{-}$ to coherently
transfer the state $\left\vert M\right\rangle =\frac{1}{\sqrt{2}}\left(e^{-i\omega_+t}\ket{+1} +e^{-i\left(\omega_-t-\left(\phi_{+}-\phi_{-}\right)\right)}\ket{-1}\right)$ to $\left\vert
0\right\rangle$, where the initial
relative phase $\phi_{+}-\phi_{-} $ is set to the same value for each round of the
experiment. Thus, the
conditional probability of measuring the state $\left\vert M\right\rangle$ is%
\begin{equation}
p_{M|H,V}\left(  t_{d}\right)  =\frac{1\pm\cos\alpha\left(  t_{d}\right)  }%
{2}, 
\label{equation2}
\end{equation}
where $\alpha\left( t_{d}\right) =\left(
\omega_{+}-\omega_{-}\right) t_{d}+\left( \phi_{+}-\phi_{-}\right) $.
Eq. \ref{equation2} 
indicates that the two conditional probabilities should oscillate with a $\pi$ phase difference
as a function of the photon detection time $t_d$. This can be understood as follows.
In the presence of Zeeman
splitting ($\delta \omega\neq0$)
the NV center's spin state is entangled with
both the polarization and frequency of the emitted photon. The photon's frequency provides which-path information about its decay.
In the spirit of quantum eraser techniques, the detection of $\ket{H}$ or $\ket{V}$ at $t_d$ with high time resolution
($\sim 300 $ ps $\ll 1/\delta\omega$) erases the frequency
information \shortcite{Scully1982,Wilk:Science2007}. When the initial relative
phase between the microwave fields $\Omega_{\pm1}$
is kept constant, the acquired phase difference $(\omega_+ - \omega_-) t_d$ gives rise to oscillations in the conditional probability and produces an effect equivalent to varying the
relative phase in the measured superposition, allowing us to
verify the coherence of the spin-photon entangled state. 

The detection times of ZPL photons are recorded during the experiment without any time gating, which allows us to study spin-photon correlations without reducing the count rate. The resulting data are analyzed in two different ways. First, we time-bin the data and use it to evaluate the conditional probabilities of measuring spin state $\ket{M}$ as a function of $\ket{H}$ or $\ket{V}$ photon detection time (Figures\ref{EntangleFig4} c,d). Off-diagonal elements of the spin-photon density matrix are evaluated from a simultaneous fit to the binned data and the time bins are chosen to minimize fit uncertainty as described in \shortcite{ToganNature2010}. The resulting conditional probabilities are used to evaluate a lower bound on the entanglement fidelity of $F\geq 0.69 \pm 0.068$, above the classical limit of 0.5, indicating the preparation of an entangled state.
 
We further reinforce our analysis using the method of maximum likelihood estimate. As described in \shortcite{ToganNature2010}, this method is applied to raw, un-binned ZPL photon detection and spin measurement data and yields a probability distribution of a lower bound on the fidelity. Consistent with the time-binned approach,  we find that our data are described by a near Gaussian probability distribution associated with a fidelity of $F \geq 0.70\pm0.070$. Significantly, the cumulative probability distribution directly shows that the measured lower bound on the fidelity is above the classical limit with a probability of 99.7 \%.  

Several experimental imperfections reduce the observed
entanglement fidelity. First, the measured strain
and magnetic field slightly mixes $\ket{A_2}$ state with the
other excited states. We estimate that $\ket{A_2}$ state imperfection and photon depolarization in the setup together reduce the fidelity by 12
\%, the latter being the dominant effect. Imperfections in readout and echo microwave pulses decrease the fidelity by 3 \%. Other error sources include finite signal
to noise in the ZPL channel (fidelity decrease 11 \%), as well as timing jitter (another 4 \%).  The resulting expected fidelity (73 \%) is consistent with our
experimental observations. Finally, the entanglement generation succeeds with
probability $p \sim 10^{-6} $, which is limited by low collection and detection
efficiency as well as the small probability of ZPL emission. 

The demonstration of spin-photon entanglement is the first step in implementing a quantum network architecture for long-distance quantum communications using NV centers \shortcite{Kimble:Nature2008}. Such an entangled state connects the solid state, spin-based local quantum register with optical photons, which can then be disseminated over free space or through optical fibers. Following the above experiment, tremendous progress has been made in achieving the next steps in building a diamond-based quantum network. Hong-Ou-Mandel interference was demonstrated with indistinguishable photons from remote NV centers hosted in separate diamond samples \shortcite{Sipahigil2012,BernienPRL2012}. Combined with spin-photon entanglement, this recently allowed for the remote entanglement of two NV centers separated by a distance of three meters through entanglement swapping \shortcite{BernienNature2013}. 

\section{Laser cooling and real-time measurement of nuclear spin 
environment of a solid-state qubit}
\label{CPTChapter}

%\subsection{Introduction}
%  Optical  cooling, continuous observation of quantum systems, and feedback control are of fundamental importance for achieving this goal.  Coherent population trapping (CPT) is a textbook example of  a quantum optical technique that is used to manipulate isolated atomic and molecular systems \shortcite{Scully97}. 
In this section, we describe the use of  the NV center's optical transitions and coherent population trapping (CPT) to sense and manipulate its surrounding spin bath, which consists of the spin of the $^{14}$N nucleus associated with the NV center itself and the $^{13}$C nuclei in the diamond lattice. Over the past two decades, CPT has been employed for laser cooling of neutral atoms and ions \shortcite{Aspect88}, creation of ultra-cold molecules \shortcite{Jin08}, optical magnetometry \shortcite{Scully92,Budker07}, and atomic clocks \shortcite{Vanier05}, as well as for slowing and stopping light pulses \shortcite{Fleischhauer:RMP2005}.  
%The same technique has been applied to solid-state systems including %the nitrogen vacancy center in diamond\cite{SantoriPRL06}, and %individual quantum dots \cite{Xu08} where interesting dynamics between %the quantum dot and its nuclear environment have been observed.
%Here we show how this quantum optical technique can be used for cooling, measurement and manipulation of artificial atom-like solid-state systems and their local environment\cite{Issler10, Xu08, Stepanenko06, Geza06}. Specifically, the method is applied to real-time projective measurement and control of the nuclear spin environment surrounding the electronic spin qubit associated with individual Nitrogen Vacancy (NV) centers in diamond \cite{Hanson06,Batalov09}.%  \cite{Petta05,Koppens06}.  
%Our experimental realization  makes use of individual Nitrogen Vacancy color centers in diamond ~ \cite{Hanson06,Batalov09}.
The electronic spin of the NV center is a promising system for extending these techniques to the solid state.
The NV center has a long-lived spin triplet as its electronic ground state \shortcite{Manson2006}%  \cite{Wrachtrup01,Jelezko04a}
, whose $m_s=\pm1, 0$ sublevels are denoted as $\ket{\pm1}$ and $\ket{0}$.
In pure samples, the electron spin dynamics are governed by interactions with the spin-1 \fortn nucleus of the NV center and spin--1/2 \carb nuclei present in  1.1~\% natural abundance in the diamond lattice (\figref{\ref{FigureCPT1}a}). 
%These interactions are now being actively  explored  % \cite{NeumannScience08}. 
Control over nuclear spins \shortcite{DuttScience2007,Neumann10}
is of interest for both fundamental 
studies and for applications such as nanoscale magnetic sensing
 \shortcite{Jero08,Gopi08} and realization of quantum networks \shortcite{ChildressScience2006,ToganNature2010}. 
%Recent advances include coherent control over individual proximal %nuclei that can be 
%isolated spectroscopically \shortcite{%Neumann08, 
%Dutt06},
%as well as single shot measurement of the 
%\fortn nuclear spin \shortcite{Neumann10}. 
Here we achieve such control via two complimentary methods: 
effective cooling of nuclear spins through nuclear state selective CPT \shortcite{Issler10} and conditional preparation based on fast
measurements of the nuclear environment and subsequent post-selection \shortcite{Geza06}.

While most prior work involved the use of microwave and RF fields for 
manipulating both the electronic and nuclear spin states, we utilize all-optical control of the electronic spin \shortcite{Robledo2010,SantoriPRL2006,Buckley10}. Specifically, we make use of $\Lambda$-type level configurations involving the NV center's $\ket{A_1}$ and $\ket{A_2}$ optically excited electronic states and the $\ket{\pm1}$ ground states (\figref{\ref{FigureCPT1}a}) \shortcite{ToganNature2010,MazeNJP2011}.  At low temperatures ($<$10 K) and in the limit of zero strain, $\ket{A_1}$ and $\ket{A_2}$ are entangled states of spin and orbital momentum
%given by $\ket{A_{2, 1}}=\frac{1}{\sqrt{2}}(\ket{E_-}\ket{+1}\pm\ket{E_+}\ket{-1}),$
%where $\ket{E_{\pm}}$ are orbital states with angular momentum projection $\pm1$ along the NV axis. These states are 
coupled to the $\ket{+1}$ ($\ket{-1}$) state with $\sigma_-$ ($\sigma+$) circularly polarized light. Correspondingly, excitation with linearly polarized light
% of the 
%form $\Omega (e^{i\phi}\sigma_-+e^{-i\phi}\sigma_+)$ 
drives the NV center into a so-called dark superposition state 
%$\ket{D_{2, 1}}=\frac{1}{\sqrt{2}}(e^{i\phi}\ket{+1}\mp e^{-i\phi}\ket{-1})$
when the two-photon detuning is zero \shortcite{Scully97}. In the present case the two-photon detuning is determined by the Zeeman splitting between the $\ket{\pm1}$ states due to the combined effect of the Overhauser field originating from the nuclear spin environment and any externally applied magnetic field \shortcite{Issler10,Stepanenko06}. When the external field exactly compensates the Overhauser field, the electronic spin of the NV center is pumped into the dark state after a few optical cycles and remains in the dark state, resulting in
vanishing fluorescence. This is the essence of the dark resonances and CPT.

\subsection{Coherent population trapping with NV centers}
\begin{figure}[htbp]
\begin{center}
\includegraphics[width = 0.9 \textwidth]{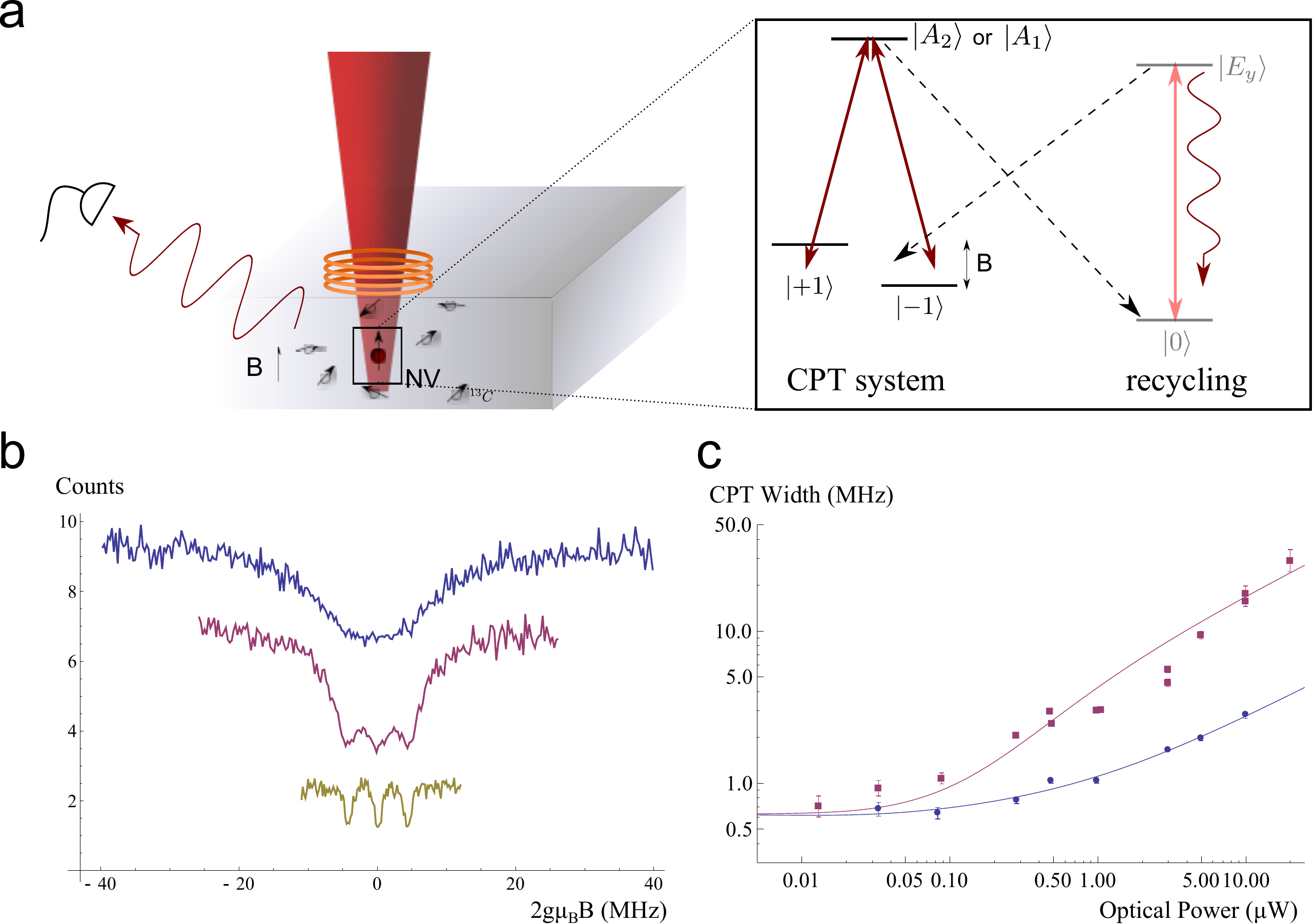}
\caption{Coherent population trapping in NV centers \textbf{a.} The $\Lambda$-type transitions between the ground states $\ket{\pm1}$ and excited states $\ket{A_{1, 2}}$ of a single NV center are addressed with a CPT laser, while a recycling laser drives the $\ket{0}$ to $\ket{E_y}$ transition. 
%The $\ket{\pm1}$ states are Zeeman shifted by 
An external magnetic field is applied using a solenoid.
%, while the nuclear environment of the NV center additionally shifts the $\ket{\pm1}$ states due to the Overhauser field. 
\textbf{b.} Photon counts from NVa in a 300 $\mu$s window are plotted versus the applied field for 10 $\mu$W (blue), 3 $\mu$W (red), and 0.1 $\mu$W (yellow) of laser power addressing the $\ket{A_2}$ state. Blue and red datasets are shifted vertically by 5 and 2 counts for clarity. \textbf{c.} Width of individual \fortn CPT lines versus CPT laser power when the $\ket{A_{1}}$ (blue) or $\ket{A_{2}}$ (red) state is used. Errorbars in all figures show $\pm1$ s.d.  Solid curves represent theoretical models. Reprinted from (Togan \textit{et al.}, 2011).}\label{FigureCPT1}
\end{center}
\end{figure}

In our experiments, the $\ket{A_{1}}$ and $\ket{A_2}$ states are
separated by approximately 3 GHz and are addressed individually with a single linearly polarized laser at near zero magnetic field.
%at $\sim$ 7 K. 
Since there is a finite branching ratio from the $m_s=\pm1$ manifold of the electron spin into the $\ket{0}$ state, we use a recycling laser that drives the transition between $\ket{0}$ and the $\ket{E_y}$ excited state, which  decays with a small but non-vanishing probability ($\sim10^{-2}$) back to the $\ket{\pm1}$ states. Figure~\ref{FigureCPT1}b presents experimental observation of  the CPT spectrum  as a function of an external magnetic field at three different  powers of a laser tuned to the $\ket{\pm1} \rightarrow \ket{A_2}$ transition. While a broad resonance is observed at high power levels,  as the power is reduced, we clearly resolve three features in the spectrum separated by $4.4$ MHz, which is two times the hyperfine splitting between three  \fortn nuclear spin states. This separation corresponds to the magnetic field required to bring the electronic $m_s=\pm1$ hyperfine states with equal nuclear spin projection ($m_I=\pm1, 0$) into two-photon resonance. 

The dependance of the CPT resonance width upon the laser power, shown in Figure~\ref{FigureCPT1}c, reveals an important role played by repumping on the near-cycling $\ket{0} \leftrightarrow \ket{E_y}$ transition.  In contrast to a conventional, closed three-level system,
this recycling transition can be used  to enhance the utility of our CPT system by both decreasing the width of the CPT resonance and increasing the signal to noise ratio. The $\ket{A_1}$ state
decays into the $m_s=0$ ground state through the singlet with a substantial probability of $\sim 40\%$. However,  the population is returned back into the  $m_s=\pm1$ state from $\ket{E_y}$ only after $\sim100$ optical excitation cycles. 
As a result, away from the two-photon resonance, the NV quickly decays to the $\ket{0}$ state after being excited, where it then scatters many photons through the $\ket{0} \leftrightarrow \ket{E_y}$ cycling transition before returning to the $\Lambda$ system. If the NV center is not in a dark state, this process effectively increases the number of photons we collect by $2/\eta = \gamma_{s1}/\gamma_{ce}$, where $\gamma_{ce}$ is the cross transition rate from  $\ket{E_y}$ into $\ket{\pm1}$ and ${\gamma_{s1}}$ is the rate from $\ket{A_1}$ to the singlet. The cycling  effect also reduces the width of the CPT line since the $\ket{0} \leftrightarrow \ket{E_y}$ transition  is quickly saturated away from two-photon resonance, provided that the CPT laser excitation rate exceeds the leakage rate out of recycling transition.    Significantly, both of these effects lead to  improved sensitivity of dark resonances to small changes in two-photon detuning.

To demonstrate this effect, the widths of dark resonances observed via
excitation of $\ket{A_1}$ and $\ket{A_2}$ are compared in Figure~\ref{FigureCPT1}c. Through an independent measurement of the branching ratios (see \shortcite{ToganNature2011}), we determined that $\ket{A_{1(2)}}$ corresponds to an open (nearly closed) 
$\Lambda$ system with $\eta_{A_1}\sim 3.1\times 10^{-2}$
%(31.3\pm0.3)\times10^{-3}$
 ($\eta_{A_2}\sim 2.6$).
 %2.607\pm0.005$. The effect of this difference between the $\ket{A_1}$ and $\ket{A_2}$ states is clearly shown in Figure~\ref{FigureCPT1}c.  
%Note, in particular, that for a case of an open $\Lambda$ transition the resonance width scales as a square root of optical power.    
These experimental 
results are compared with  a theoretical model described in \shortcite{ToganNature2011}, which predicts that the resonance linewidth $\delta_0$ is given by
$\delta_0 =  \sqrt{R_{A}^2/\left[1+\frac{1}{\eta}\left(\frac{R_{A}}{R_{E}}+\frac{2R_{A}}{\gamma}\right)\right]}\sim \sqrt{R_A R_E \eta\gamma/(R_E+\gamma)}$ for small $\eta$, where $R_{A(E)}$ corresponds to the optical excitation rate by a laser tuned to the $A(E)$ state and $\gamma$ is the decay rate of excited states. The width at low powers is determined by the random magnetic field associated with surrounding \carb nuclear states. When this line broadening mechanism is taken into account,
the experimental results are 
in excellent agreement with these predictions, plotted as solid lines, showing that a high degree of optical control 
over the electronic spins of NV center can be achieved.

Subsequent experiments \shortcite{ToganNature2011} showed that CPT can be used for not only probing the nuclear spin environment of the NV center, but also controlling and preparing the state of the spin bath. For the \fortn spins, it is possible to optically cool the nuclear spin states using dark resonances. The physics is reminiscent of laser cooling of atomic motion via 
velocity-selective CPT \shortcite{Aspect88,Issler10}. 
 A redistribution of the \fortn spin state population upon optical excitation takes place because the hyperfine coupling in the excited electronic state of the NV center is enhanced by a factor of $\sim20$ compared to the ground state \shortcite{Fuchs:PRL2008}. If the external field is set such that, for example, the $m_I=0$ hyperfine states are in two photon resonance, only the states with nuclear configuration $m_I=\pm1$  will be promoted to the excited states, where flip-flops with the electron spin will change the nuclear spin state to $m_I=0$. When the NV center spontaneously decays into the dark superposition of electronic spin states, optical excitation will cease, resulting in effective polarization (cooling) of nuclear spin into $m_I=0$ state. 
 
\begin{figure}[htbp]
\begin{center}
\includegraphics[width = 0.6 \textwidth]{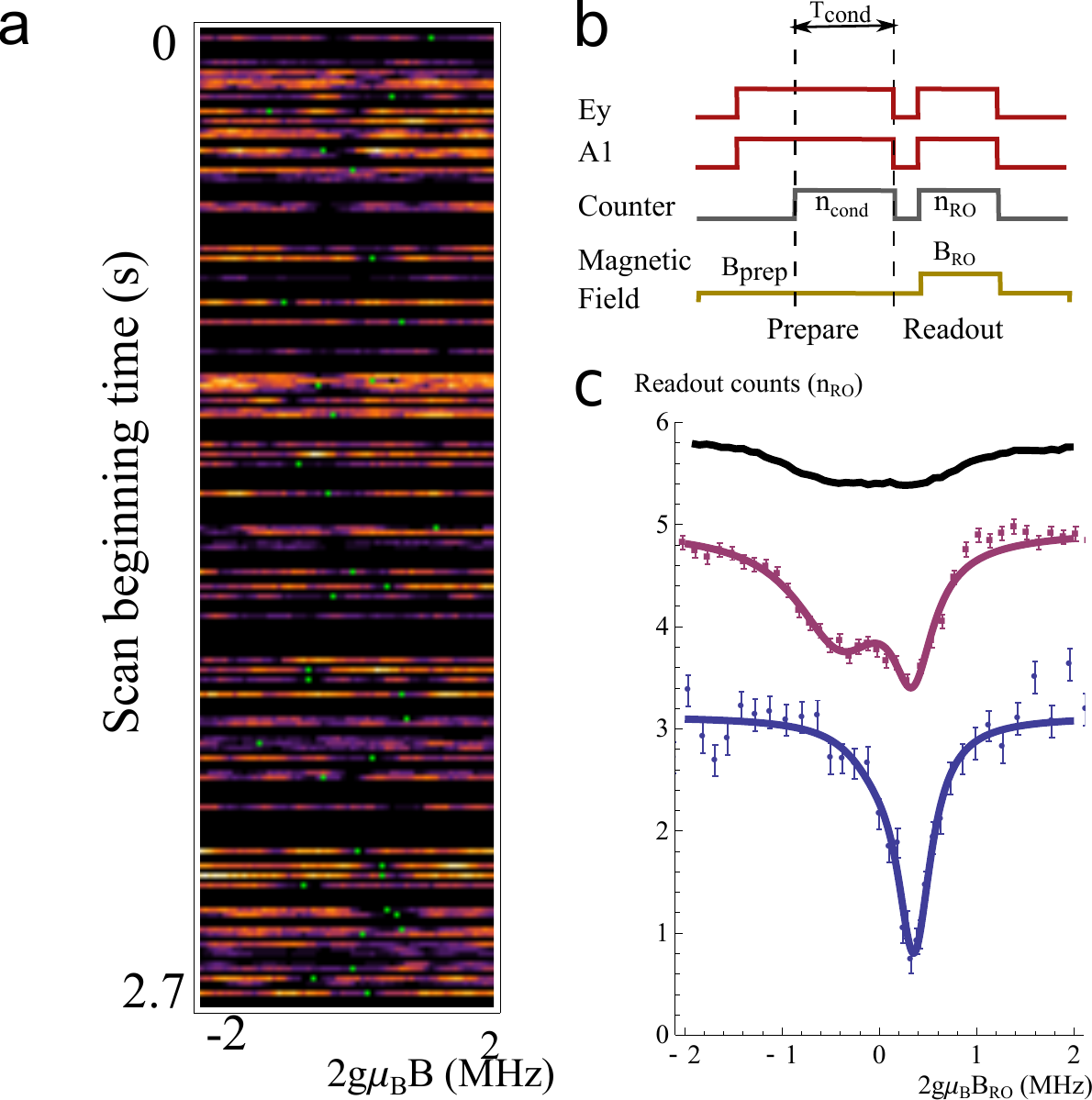}
\caption{Fast measurement and preparation of \carb spin bath. \textbf{a.} Counts from 200 successive fast magnetic field ramps are shown on horizontal lines. The centers of Lorentzian fits to individual runs are indicated with a dot. \textbf{b.} Pulse sequence for preparation and subsequent measurement of the \carb configuration. $B_{\mathrm{prep}}$ is set within the central \fortn line while  $B_{\mathrm{RO}}$ is varied to cover all associated \carb states. Counts during a conditioning window of length $T_{\mathrm{cond}}$ ($n_{\mathrm{cond}}$) at the end of preparation and the readout window ($n_{\mathrm{RO}}$) are recorded for each run.  Data presented is an average of many such experimental runs. \textbf{b.} $n_{\mathrm{RO}}$ versus $B_{\mathrm{RO}}$ is shown in the middle plot with double Lorentzian fit. The same dataset analyzed by keeping only events with $n_{\mathrm{cond}}=0$ is shown in the bottom plot. Unprepared \carb distribution is shown in the top plot for comparison (shifted by 4.3 counts for clarity). Reprinted from (Togan \textit{et al.}, 2011).}\label{FigureCPT2}
\end{center}
\end{figure}
 
We also extended this technique to control the many-body environment of the NV center, consisting of \carb nuclei distributed throughout the diamond lattice. The large number of  nuclear spin configurations associated with an unpolarized environment  results in a random Overhasuer field ($B_{\mathrm{ov}}$) with unresolved hyperfine lines. It produces a finite CPT linewidth in  measurements that average over all configurations of the \carb spin bath (Figure~\ref{FigureCPT1}b, c).  
However, one can make use of fast measurements and the long correlation time  ($T_1^{nuc}$) associated with evolution of the nuclear bath to observe its instantaneous state and and manipulate its dynamics. For example, such a fast measurement is illustrated in Figure~\ref{FigureCPT2}a, where the externally applied field is ramped across a single \fortn $m_I = 0$ line while the CPT lasers are on, and counts collected in 80 $\mu$s time bins are plotted for successive individual runs, many of which distinctly show a narrow dark region.
 Lorentzian fits to selected experimental scans reveal ``instantaneous" CPT resonances with linewidths that are over a factor three less than those of the averaged measurement. The motion of the  dark line centers (green dots in Figure~\ref{FigureCPT2}a) indicates that the instantaneous field evolves in time. Fast measurements can also be used to conditionally prepare the \carb environment of the NV center in a desired state with post-selection. 
Here, the nuclear spins are first prepared though optical excitation at a fixed value of the external magnetic field $B_{\mathrm{prep}}$, as shown in Figure~\ref{FigureCPT2}b. By conditionally selecting zero photon detection events
during this preparation step, we can select the states of the \carb
environment  with vanishing two photon detuning
$\delta =2g\mu_B (B_{\mathrm{prep}} + B_{\mathrm{ov}}) =0$,  where $\mu_B$ is the Bohr magneton. The resulting modified distribution of Overhauser field is then probed in a readout step by rapidly the changing magnetic field value to $B_{\mathrm{RO}}$, which is scanned across the resonance.  The middle curve in Figure~\ref{FigureCPT2}c shows (unconditioned) readout counts recorded following the preparation step, while 
the bottom curve shows the result of 
measurement-based preparation.
The measured width of such a conditionally prepared distribution is significantly smaller than the width corresponding to individual \fortn resonances obtained without preparation.

%
%\begin{figure}[htbp]
%\begin{center}
%\includegraphics[width = 0.9 \textwidth]{FigureCPT2}
%\caption{Fast measurement and preparation of \carb spin bath. \textbf{a.} Counts from 200 successive runs are shown on horizontal lines. The centers of Lorentzian fits to individual runs are indicated with a green dot. \textbf{b.} Pulse sequence for preparation and subsequent measurement of the \carb configuration. $B_{\mathrm{prep}}$ is set within the central \fortn line while  $B_{\mathrm{RO}}$ is varied to cover all associated \carb states. A preceding green laser pulse and \fortn optical pumping step with the $A_2$ and $E_y$ lasers are not shown. Counts during a conditioning window of length $T_{\mathrm{cond}}$ ($n_{\mathrm{cond}}$) at the end of preparation and the readout window ($n_{\mathrm{RO}}$) are recorded for each run.  Data presented is an average of many such experimental runs. \textbf{b.} $n_{\mathrm{RO}}$ for NVb versus $B_{\mathrm{RO}}$ is shown in red with double Lorentzian fit. The same dataset analyzed by keeping only events with $n_{\mathrm{cond}}=0$ is shown in blue. Unprepared \carb distribution is shown in black for comparison (shifted by 4.3 counts for clarity).}
%\label{FigureCPT2}
%\end{center}

\section{Coherent optical transitions in implanted nitrogen vacancy centers}
\label{ImplantationChapter}

Since one-of-a-kind natural diamond samples such as the one investigated in the previous sections cannot be used as reproducible starting materials for scalable systems, recent experiments have made use of NV centers incorporated into high quality synthetic diamond during the growth process
%. The extrinsically broadened linewidth observed in these bulk diamond samples have typically been on the order of ten times the lifetime limited linewidth 
\shortcite{BernienPRL2012,AcostaPRL2012,StaceyAM2012}. 
%Such a large frequency uncertainty is detrimental for almost every application that involves coherent coupling between NV centers and optical fields. 
%For example, recent experiments demonstrating long distance NV-NV entanglement relied on interference between photons from two NV centers that are identical in frequency \shortcite{BernienNature2013}. While the frequency uncertainty due to spectral diffusion can be partly alleviated by increasing the timing resolution of the detectors, dynamic feedback techniques, or post-selection, these methods generally decrease the overall efficiency. 
In addition to imperfections in their optical coherence, NV centers introduced during the diamond growth process are limited in their usefulness due to their low concentration and random locations. For example, in order to incorporate NV centers into nanophotonic  or nanomechanical devices, it is necessary to develop a method to create a controllable concentration of emitters in a well defined device layer while maintaining their excellent optical properties. 

This section describes how NV centers can be created in a surface layer that exhibit coherent optical transitions with nearly lifetime-limited linewidths even under the effects of spectral diffusion. We accomplish this by first introducing NV centers at a well defined and controllable depth using ion implantation \shortcite{PezzagnaNJP2013}. We then demonstrate how to suppress the fundamental cause of spectral diffusion by creating a diamond environment that is nearly free of defects that contribute to charge fluctuations. This is achieved through a combination of annealing and surface treatments. Finally, we further improve the optical properties of the NV centers by combining our approach with a newly developed technique for preventing photoionization of the remaining charge traps \shortcite{SiyushevPRL2013}.

\begin{figure}[ht]
\begin{center}
\includegraphics[width=0.9\textwidth]{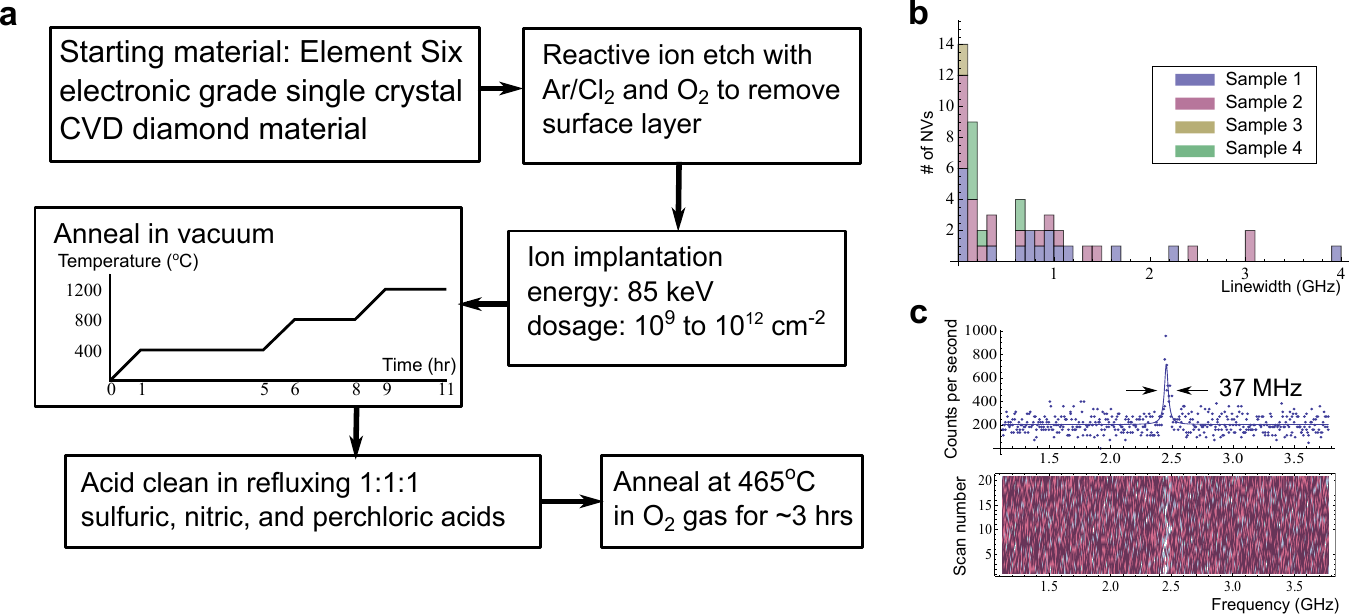}
\caption{Procedure for creating spectrally stable NV centers through ion implantation, annealing, and surface treatment. The Ar/Cl$_2$ (O$_2$) etch was done using an ICP power of 400 W (700 W), a RF power of 250 W (100 W), a DC bias of 423 V (170 V), a chamber pressure of 8 mTorr (10 mTorr), and a gas flow rate of 25/40 sccm (30 sccm). The sample temperature was set to 17$^\circ$C. Reprinted from (Chu \textit{et al.} 2014).}
\label{fig:SamplePrep}
\end{center}
\end{figure}

Fig. \ref{fig:SamplePrep}(a) summarizes the experimental procedure to create shallow implanted NV centers with narrow optical lines. As our starting material we use single crystal diamond from Element Six grown using microwave assisted chemical vapor deposition (CVD). The single crystal samples were homoepitaxial grown on specially prepared $\langle$100$\rangle$ oriented synthetic diamond substrates, taking care to ensure that the surface quality of the substrates was as high as possible to reduce sources of dislocations. The CVD synthesis was performed using conditions as described in \shortcite{Isberg2002,Markham2011}. The resultant diamond material is measured using electron paramagnetic resonance (EPR) to have a nitrogen concentration of less than 5 parts per billion. The surface layer of these samples is generally damaged and highly strained because of mechanical polishing. Therefore, we first remove the top several micrometers of the sample using a 30 min Ar/Cl$_2$ etch followed by a 20 min O$_2$ etch in an UNAXIS Shuttleline inductively coupled plasma-reactive ion etcher. We then implant $^{15}$N$^+$ ions at an energy of 85 keV at a variety of doses from $10^9$/cm$^{2}$ to $10^{12}$/cm$^{2}$. According to simulations done using Stopping and Range of Ions in Matter (SRIM), this should result in a mean nitrogen stopping depth of $\sim$100 nm with a straggle of $\sim$20 nm \shortcite{SRIM}. Subsequently, we anneal the implanted samples under high vacuum ($P < 10^{-6}$ Torr). Our standard annealing recipe consists of a 4 hour 400$^\circ$C step, a 2 hour 800$^\circ$C step, and a 2 hour 1200$^\circ$C step, with a 1 hour ramp up to each temperature. The motivation behind these temperature steps will be explained later in the text. It is important that a good vacuum is maintained throughout the annealing process, since diamond etches and graphitizes under residual oxidizing gasses at such high temperatures \shortcite{SealPSS1963}. Following the anneal, we remove graphitic carbon and other surface contaminants by cleaning the diamonds for one hour in a 1:1:1 refluxing mixture of sulfuric, nitric, and perchloric acids. Finally, we perform a 465$^\circ$C anneal in an O$_2$ atmosphere in a rapid thermal processor (Modular Process Technology, RTP-600xp) in three 48 minute steps following the recipe used in  \shortcite{FuAPL2010}. This oxygen anneal is believed to remove $sp^2$ hybridized carbon and result in a more perfect oxygen termination of the surface than acid cleaning alone. We have observed that it enhances charge stability in the NV$^-$ state and helps to reduce blinking and photobleaching. 

Let us focus on samples implanted with a $^{15}$N$^+$ ion dose of $10^{10}$/cm$^{2}$. Fig. \ref{fig:SamplePrep}(b) shows a histogram of linewidths for 50 NV centers in four such samples. While there is a range of linewidths, a majority of NV centers exhibit linewidths less than 500 MHz, which is comparable to naturally occurring NV centers deep inside electronic grade CVD diamond \shortcite{BernienPRL2012}. This is an indication that we have eliminated most impurities with fluctuating charge states in the vicinity of the NV centers, which may include both impurities on the nearby surface and defects inside the diamond introduced by the implantation process. In Fig. \ref{fig:SamplePrep}(c), we present an example spectrum of the narrowest linewidth we have observed, along with the results of twenty-one successive PLE scans to demonstrate the long-term stability of the transition line. Our measured linewidth of $37\pm4.8$ MHz under conventional 532 nm repumping is, to the best of our knowledge, a record for the extrinsically broadened linewidth of NV centers. 

\section{Diamond-based nanophotonic devices for quantum optics}
\label{NanophotonicsChapter}

\subsection{Cavity QED with NV centers: the idea}

\begin{figure}
\centering
\includegraphics[width=\textwidth]{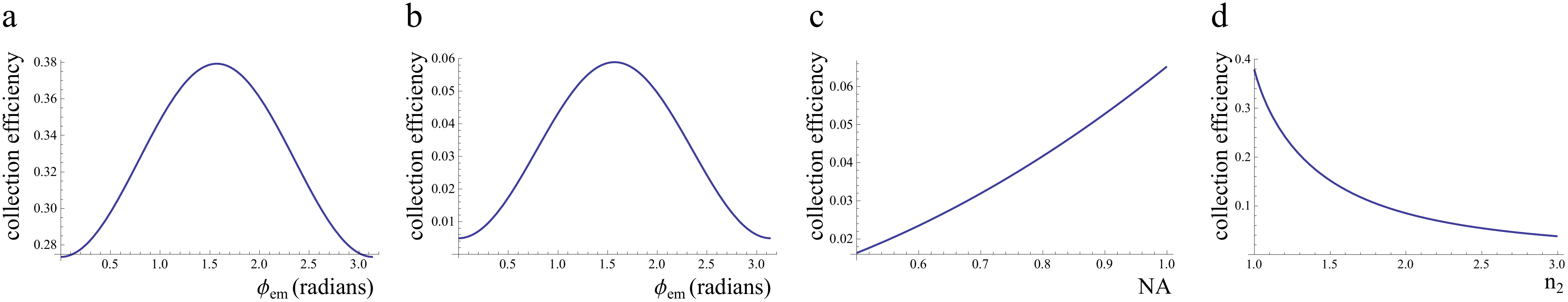}
\caption{Collection efficiency of light from a dipole emitter. \textbf{a.} Collection efficiency vs. dipole orientation for NA=0.95 and $n_2$=1. \textbf{b.} Collection efficiency vs. dipole orientation for NA=0.95 and $n_2$=2.4. \textbf{c.} Collection efficiency vs. NA for $\phi_{em}=\pi/2$ and $n_2=2.4$. \textbf{d.} Collection efficiency vs. $n_2$ for $\phi_{em}=\pi/2$ and NA=0.95.}
\label{f:collectionEfficiency}
\end{figure}

In the previous sections, we have demonstrated the potential of using the NV center's optical properties for a variety of different applications. However, the NV center is not a perfect optical emitter for several reasons. First, the Debye-Waller factor $\xi$, defined as the ratio of emission into the ZPL vs. the total emission, is only 3-5 $\%$. This is dictated by the relatively large shift in nuclear coordinates between the NV center's ground and excited states, which, by the Franck-Condon principle, results in decay from the excited state into phonon-coupled higher vibrational levels of the ground state \shortcite{GaliNJP2011}. In addition to such intrinsic properties of the NV center, there are also external effects such as spectral diffusion that result in non-ideal optical properties. While we can eliminate these effects to a large degree, as demonstrated in Chapter \ref{ImplantationChapter}, some residual imperfections in spectral stability usually remain. Additional techniques to stabilize the transition frequency, such as feedback techniques with applied electric fields, reduce the repetition rate of the experiments. Finally, it is generally difficult to extract photons from bulk diamond material with high efficiency. Due to the high index of refraction, total internal reflection confines most of the light inside the diamond at an air interface. In general, as shown in \shortcite{PlakhotnikOpCom1995}, the collection efficiency of a lens or objective with numerical aperture NA in a medium with index $n_1$ for light from a dipole source in index $n_2$ is given by
\begin{equation}
C=\frac{1}{8}\left(4-3\textrm{cos}\theta_{max}-\textrm{cos}^3\theta_{max}+3(\textrm{cos}^3\theta_{max}-\textrm{cos}\theta_{max})\textrm{cos}^2\phi_{em}\right),
\end{equation}
where $\phi_{em}$ is the angle of the dipole axis with the optical axis, and $\theta_{max}=\textrm{sin}^{-1}(\textrm{NA}*n_1/n_2)$ is the effective NA of the collection optics. For reference, the collection efficiency is plotted against several parameters in Figure \ref{f:collectionEfficiency}. As can be seen, even for optimal dipole orientation and a 0.95 NA objective in air, only around $6\%$ of the emitted photons are collected from a NV center in bulk diamond.

The combination of these factors result in a very low rate of indistinguishable, coherent photons from single NV centers. One promising path toward alleviating all of these imperfections is incorporating NV centers with photonic structures. For example, optical cavities can be used to enhance the emission into the ZPL and overcome spectral diffusion, while coupling to waveguide structures can increase the efficiency of photon collection and detection. 

The idea of using cavity QED to improve the emission properties of a NV center is as follows: When a NV center is placed inside the an optical cavity, its emission into the cavity mode is enhanced. By tuning the cavity mode to the NV center's ZPL, the ZPL emission is therefore enhanced relative to the PSB emission. When this enhancement becomes larger than the Debye-Waller factor, the ZPL becomes the dominant emission channel of the NV center. This enhancement of the emission also means that the lifetime of the NV center is decreased and the lifetime-limited linewidth of the ZPL is increased. When the lifetime-limited linewidth becomes much larger than the spectral diffusion, the NV center can be viewed as a perfect emitter of indistinguishable photons. In other words, the energy uncertainty of the excited state caused by fast decay into the cavity mode is large enough that the energy uncertainty introduced by spectral jumps is negligible. 

We can calculate lifetime of a NV center coupled to the cavity ($\tau$) compared to one that is decoupled from the cavity ($\tau_0$). They are related through a quantity called the Purcell factor, given by
\begin{equation}
P=\frac{3}{4\pi^2}\left(\frac{\lambda_0}{n}\right)^3\frac{Q}{V}\frac{|\vec{E}_{NV}\cdot\vec{\mu}_{ZPL}|^2}{|\vec{E}_{max}|^2|\vec{\mu}_{NV}|^2}=\frac{\tau_0}{\tau}-1
\label{e:purcell}
\end{equation}
where $\lambda_0/n$ is the wavelength of the resonant mode inside the cavity and Q is the quality factor of the cavity. $\vec{E}_{NV}$ and $\vec{E}_{max}$ are the electric fields at the NV and maximum field inside the cavity, respectively. The mode volume of the cavity is defined as 
\begin{equation}
V=\frac{\int |\vec{E}(\vec{x})|^2 dV}{|\vec{E}_{max}|^2}
\end{equation}
Finally, $\vec{\mu}_{ZPL}$ is the effective dipole moment of the ZPL transition, which is the only part coupled to the cavity, while $\vec{\mu}_{NV}$ is the total dipole moment of the NV center. The Debye-Waller factor is then given by $\xi=|\vec{\mu}_{ZPL}|^2/|\vec{\mu}_{NV}|^2$. The Purcell factor gives the enhancement of emission into the cavity mode over the emission when the NV is decoupled from the cavity. Here, we define ``decoupled" as an NV that is still spatially located in the cavity, but detuned from the cavity mode in frequency, as is the case in our experiments. This means that $n$ in the expression above is the effective index including all possible modes of the structure. Although this quantity can be difficult to estimate, it is somewhere between 1 and 2.4, the index of bulk diamond.

While the Purcell factor expresses the relevant enhancement of emission in terms of experimentally relevant quantities, it is equivalent to the cooperativity, which gives more physical intuition as simply the ratio between the coupling strength of the emitter to the cavity and the decay rates of the system. The Purcell factor or cooperativity can also be written as
\begin{equation}
P=\frac{2g^2}{\kappa\gamma}
\label{e:cooperativity}
\end{equation}
where $\kappa=\omega/Q$ is the decay rate from the cavity, with $\omega$ being the cavity frequency. $\gamma$ is the decay rate of the NV center into modes other than the resonant cavity mode, and is given by
\begin{equation}
\gamma=\frac{\omega^3|\mu_{NV}|^2}{3\pi\epsilon\hbar c^3}=\frac{8\pi^2|\mu_{NV}|^2}{3\epsilon\hbar}\left(\frac{n}{\lambda_0}\right)^3.
\end{equation}
The single-photon Rabi frequency $g$ is given by $g=\vec{E}_{NV}\cdot\vec{\mu}_{ZPL}/\hbar$. We emphasize here that since the cavity mode only couples to the ZPL, we use the effective dipole moment of that transition only. Inserting these expressions into Equation \ref{e:cooperativity} and using the fact that $\epsilon \int |\vec{E}(\vec{x})|^2 dV=\hbar\omega$, we recover the expression for the Purcell factor given in Equation \ref{e:purcell}.

We note here that there are several alternative definitions of the Purcell factor that might also relevant here. One can also consider the lifetime change relative to emission into bulk diamond or free space, in which case the index $n$ in Equation \ref{e:purcell} should be 2.4 or 1, respectively. Another quantity is the enhancement of the ZPL emission into the cavity relative to the ZPL emission when the NV is decoupled from the cavity. This is given by $P'=P/\xi$, and a useful quantity when comparing the Purcell enhancement to other cavity QED systems where the emitter has a single decay channel. In other words, it gives a better comparison for the quality of the emitter-cavity coupling without factoring in the internal imperfections of the emitter itself, and is the quantity usually quoted in the literature. It is also the relevant quantity for estimating the enhancement of the ZPL photon rate from the cavity compared to, say, bulk diamond. 

As explained above, in order to radiatively broaden the NV center through Purcell enhancement to overcome spectral diffusion, we require that $P\gg\frac{\Gamma}{\gamma}$, or $P'\gg\frac{1}{\xi}\frac{\Gamma}{\gamma}$, where $\Gamma$ is the extrinsically broadened linewidth. From Equation \ref{e:purcell}, we see that to obtain a large Purcell factor, we need large $Q$ and small mode volume. This can be achieved using photonic crystal cavities, where the mode volume can be designed to be $\sim(\lambda/n)^3$. Then, for a NV center with $\Gamma\sim100$ MHz, a cavity with a Q of $\sim3000$ is required. We now discuss some directions for the physical implementation of such cavities in diamond. 

\subsection{Diamond-based nanophotonic devices}

\begin{figure}
\centering
\includegraphics[width=\textwidth]{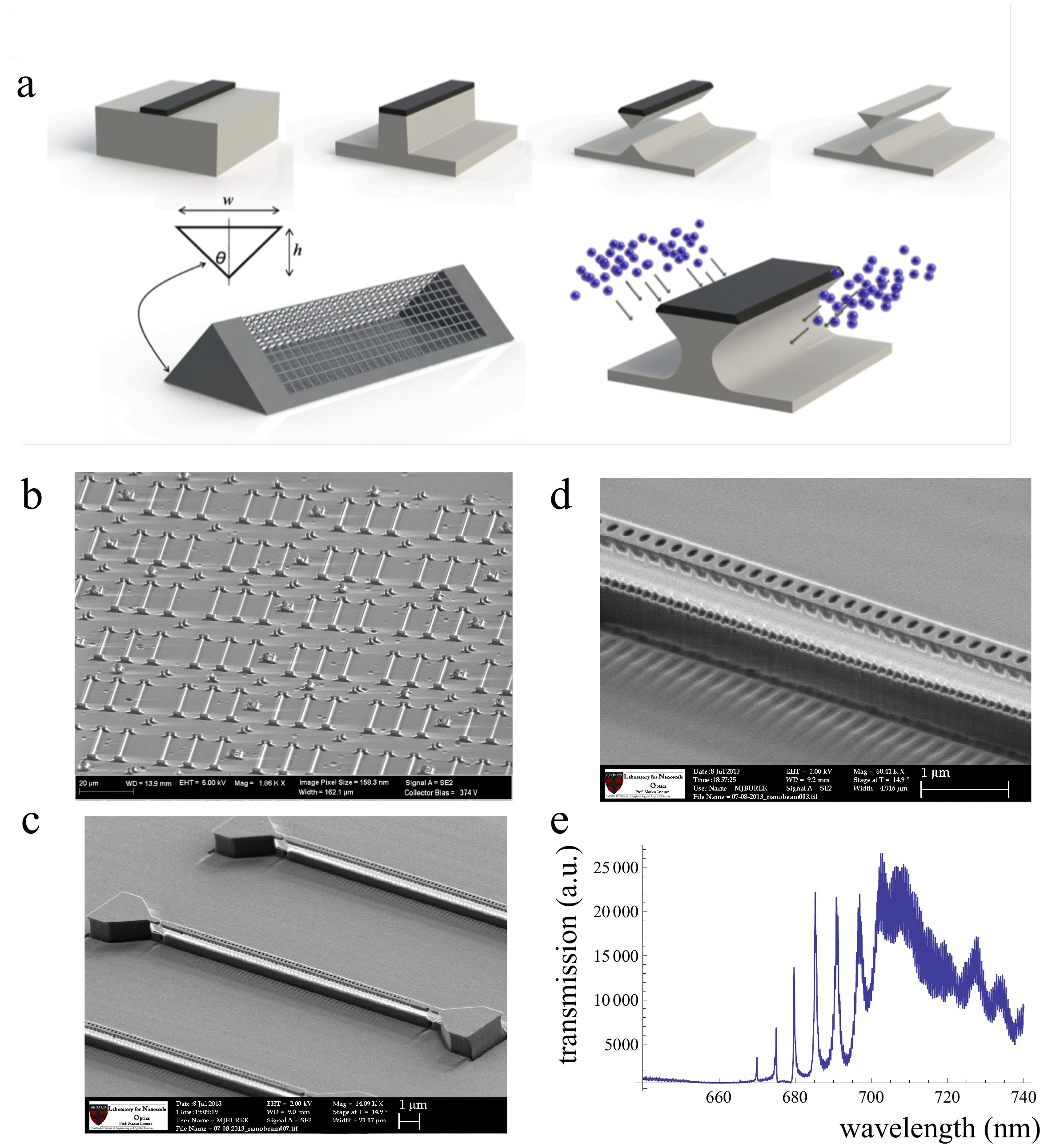}
\caption{Suspended diamond-based nanophotonic structures. \textbf{a.} Procedure for angled RIE etching. A mask is defined using EBL on the surface of the diamond. A top-down etching step performed, followed by an angle-etching step where the diamond is placed in the center of a machined aluminum faraday cage shown on the bottom left. The mask is removed, leaving a suspended structure attached at both ends to anchors (not shown) \textbf{b.} SEM image of a large array of $\sim$1000 angle-etched waveguides. \textbf{c-d.} SEM images of photonic crystal cavities formed in 1D waveguides. \textbf{e}. Transmission spectrum of one such device. See (Burek \textit{et al.}, 2012) for details.}
\label{f:angleEtch}
\end{figure}

Over the past several decades, a large range of techniques have been developed for the fabrication of nanoscale devices in semiconductor materials ranging from silicon-based materials to III-V compounds such as gallium arsenide \shortcite{EnglundNature2007,BlancoNature2000}. While many of these techniques can be transferred to the fabrication of diamond, one major difference between diamond and most semiconductor materials presents a significant challenge to the fabrication of nanoscale devices for confining light. Because of the large lattice mismatch of diamond with most other crystalline materials, it is not possible to grow high quality thin films of single crystal diamond on top of a different substrate. This means that we cannot take advantage of the index mismatch between the device and substrate to confine light through total internal reflection. Here, we will briefly describe two methods for creating diamond-based nanophotonic devices by either creating a thin film that is placed on a substrate, or directly fabricating suspended structures on the surface of a bulk diamond sample. 
The first method for fabricating diamond-based photonic crystal cavities begins with a mechanically polished diamond membrane that is thinned down using a deep reactive ion etch (RIE) in inductively coupled plasma (ICP) of oxygen to the device thickness \shortcite{HausmannNanoLett2013,Faraon2011}. Devices are then defined on these thin films by e-beam lithography (EBL) using hydrogen silsesquioxane (HSQ) resist, which forms a mask that is transferred into the diamond film during another RIE etching step. 
%Finally, the diamond structures are suspended by an isotropic RIE step that undercuts the Si substrate beneath the devices. 
%A SEM image of 1D photonic crystals produced using this method is shown in Figure \ref{f:cavities}a. 
%Using such devices, we have observed enhancement of the ZPL fluorescence by a factor of 7. This work is described in detail in \shortcite{HausmannNanoLett2013}, and similar techniques have been used to create diamond photonic crystal structures in \shortcite{Faraon2011}. 
%The method described above has the drawback that, since the membranes are mechanically polished, there is often a gradient in the thickness, which transferrers to the thin film. This means that only a small region of the devices that are made using the film has the desired thickness and wavelength. In addition, the thin films are difficult to handle and post-process. 
The second technique for scalable fabrication of nanophotonic structures involves undercut structures suspended above the surface of bulk diamond. This is achieved using a technique described in \shortcite{BurekNL2012}, where the diamond is placed inside a faraday cage during RIE so that the ions are redirected and impinge on the surface at an angle. As a result, instead of etching the diamond in the direction perpendicular to the surface, it undercuts the material beneath the mask from two directions, resulting in suspended structures with a triangular cross-section that are suspended from or supported by anchors. 
As shown in \shortcite{BurekNL2012} and Figure \ref{f:angleEtch}, we can make a variety of devices including ring resonators, cantilevers, and 1D photonic crystal cavities using this method. In addition, 1D waveguide structures can be incorporated into hybrid cavities where a Bragg structure is defined by another material around an NV inside the waveguide, such as PMMA.
%We have fabricated 1D photonic crystal cavities using this method by etching an array of holes with tapered aspect ratios along the waveguide, again using HSQ as an EBL defined mask, as shown in Figure \ref{f:angleEtch}. We can characterize these cavities using by coupling a broadband light source into the waveguide through a notch at one end and collecting the transmitted light that is scattered from a notch at the other end. An example of such a transmission spectrum is shown in Figure \ref{f:angleEtch}. Cavity resonances can be seen near the dielectric band edge with Q's of more than 3000. 

\subsection{Example: Purcell enhancement from a coupled NV-cavity system}

As an example, we present here a measurement demonstrating the Purcell enhancement of a NV center's ZPL emission by a nanoscale hybrid diamond-PMMA cavity \shortcite{deleon2014}. These measurements were made on a NV center with a 3 GHz linewidth inside a hybrid cavity with a Q of $\sim$600. 

In order to tune the cavity mode frequency, we use a combination of two techniques. The first method uses a O$_2$ plasma etcher (Technics, Model 220) to remove a layer of material from the structure \shortcite{HausmannNanoLett2013}. This tunes the cavity resonance to lower wavelengths because more of the cavity field resides in air than before and therefore has a higher energy. We have been able to achieve a wavelength shift of $\sim$15 nm without significant degradation in the Q factor using this method. Complementary to this blue tuning process, we can also red tune the cavity resonance by depositing neon gas on the structure when it has been cooled down inside the cryostat \shortcite{MosorAPL2005}. We note here that gas deposition is a common method for tuning photonic crystal cavities, and a tuning range of $\sim$23 nm was observed in \shortcite{HausmannNanoLett2013}. However, it is somewhat less effective for our hybrid structures since the high aspect ratio of the PMMA slabs prevent the diffusion of gas onto the diamond waveguide, where the field is concentrated. Despite this, we have occasionally been able to obtain tuning ranges of $\sim$6 nm. 

\begin{figure}
\centering
\includegraphics[width=\textwidth]{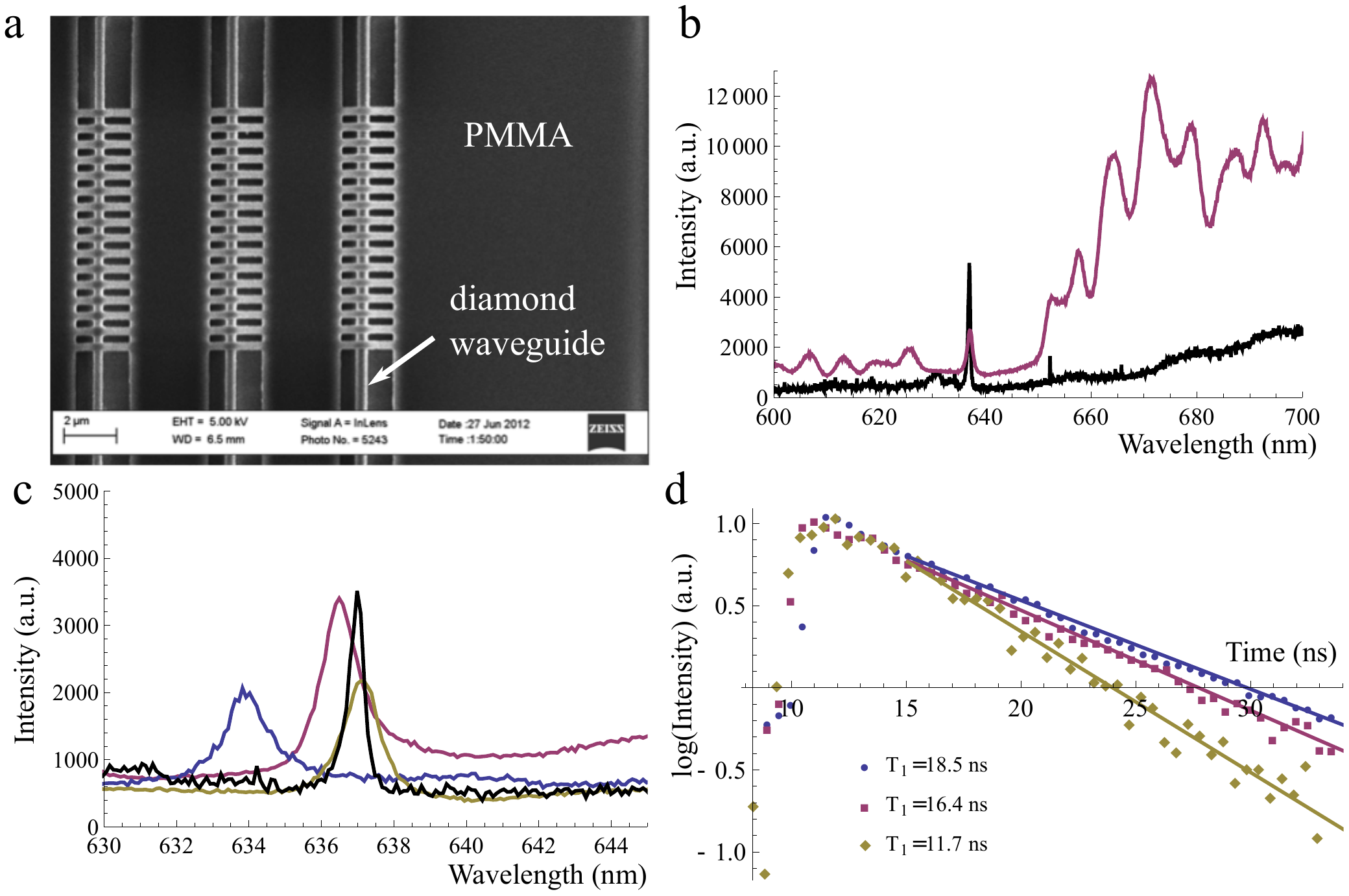}
\caption{Purcell enhancement of the NV center ZPL emission using hybrid photonic crystal cavities. \textbf{a.} SEM image of hybrid diamond-PMMA cavities. Visible are the top surface of the suspended triangular beams, the top of the the PMMA layer and slabs, along with the exposed diamond surface underneath. \textbf{b.} Transmission spectrum of hybrid cavity (magenta) and emission spectrum of NV center (black). The spectra have been scaled relative to each other to fit on the same scale. \textbf{c.} Zoomed in spectra showing ZPL of NV center (black) and cavity resonance tuned to three different frequencies (blue, magenta, and yellow). \textbf{d.} Lifetime measurements corresponding to the detunings in c., taken by exciting the NV center with a pulsed laser and collecting the PSB emission. See (de Leon \textit{et al.}, 2014)}
\label{f:purcell}
\end{figure}

To measure the change in lifetime when the NV center becomes coupled with the cavity, we first tune the cavity resonance using plasma etching until it is to the blue of the NV center's ZPL. We then gradually red tune the cavity onto resonance with the ZPL while measuring the lifetime using a filtered supercontinuum laser (NKT photonics, SuperK EXTREME) as a pulsed laser source around 532 nm. As shown in Figure \ref{f:purcell}, the lifetime decreases from 18.5 ns when the cavity is completely off resonant to 11.6 ns when the cavity is tuned to the ZPL. This gives a Purcell factor of $P=0.59$, or a ZPL enhancement of $P'=20$ for $\xi=0.03$. We emphasize here that this enhancement factor implies that the ZPL emission is now $\sim$40$\%$ of the total emission from the NV center. 
%However, the lifetime limited linewidth is almost unchanged, and a much larger Purcell factor is needed to overcome the spectral diffusion, which in this case is a factor of $\Gamma/\gamma\sim200$. 

\section{Outlook}
Following the experiments presented in Section \ref{EntanglementChapter}, there has been tremendous progress in using optical photons to connect remote NV centers. Future work will explore the possibility of combining the few-qubit local spin register with the resource of remote entanglement through, for example, teleportation of quantum information between remote, long-lived nuclear spin memories. Distribution of quantum information by teleportation has been demonstrated in several different physical systems \shortcite{NollekePRL2013,SteffenNature2013,KrauterNatPhys2013}. However, remote teleportation over long distances is an outstanding challenge with solid state systems, which, in the long run, offers the possibility of implementing more scalable local quantum nodes. In particular, one major factor in determining the ultimate bandwidth and fidelity of a remote quantum channel is the efficiency with which single photons can be made to interact with the stationary qubits and then extracted into, for example, a optical fiber. Current photon collection efficiencies and entanglement rates may be sufficient for proof-of-principle experiments, but certainly must be greatly improved for the implementation of practical quantum networks. The work described in Sections \ref{ImplantationChapter} and \ref{NanophotonicsChapter} begins to address this issue. The spectral stability of NV centers is an important factor determining the repetition rate of experiments requiring indistinguishable optical photons, such as remote entanglement and teleportation. The availability of NV centers with lifetime-limited linewidths would eliminate the need for active stabilization and post-selection of the the transition frequency in future experiments. Cavity QED has been used to enhance the interaction of photons with atomic systems and improve their collection efficiency, leading to demonstrations of quantum teleportation with high fidelity and success rates \shortcite{NollekePRL2013}. Diamond-based nanophotonics offers the possibility of achieving efficient photonic quantum devices in a solid state system. 

Finally, future quantum technologies will most likely be implemented with a combination of several types of physical systems, each with its own strengths and weaknesses. There are already many theoretical proposals and experimental demonstrations of interfacing NV centers with other quantum devices \shortcite{KuboPRL2011,RablNatPhys2010}. Most of these demonstrations make use of the NV center's unique spin properties, effectively turning its interaction with the magnetic environment from a detriment into a resource. Analogous to this, an potential direction for future exploration would be to take advantage of the NV center's optical transitions to interface it with other systems. 
%Thus far, we have treated the strain and electric field sensitivity of the NV center's excited states as a detrimental effect. However, these properties can also be used to, for example, couple multiple nearby NV centers for the purpose of investigating interesting collective effects or scaling up the local quantum register. In addition, they allow for interactions between the NV center and other structures that can modify its local strain or electric field, such as mechanical oscillators and superconducting circuits. The strength of these interactions can potentially be much stronger than the magnetic coupling using the ground state of the NV center. 
Combined with the ability to communicate information with remote systems through optical photons, the NV center becomes a quantum system that not only has uniquely attractive properties as an isolated system, but can also be made to interact with other systems in a controlled manner. Therefore, the ideas and techniques described in this chapter will likely enable researchers to use the optical properties of the NV center as a powerful resource for building complex and useful quantum technologies.

\section{Acknowledgements}

The work described in this chapter was carried out by several generations of Harvard quantum optics group members, including Lillian Childress, M. V. Gurudev Dutt, Ruffin Evans, Liang Jiang, Nathalie de Leon, Jeronimo Maze, Brendan Shields, Emre Togan, Alexei Trifonov, and Alexander Zibrov. We additionally acknowledge many discussions and collaborations with Michael Burek, Birgit Hausmann, Atac Imamoglu, Fedor Jelezko, Marko Loncar, Philip Hemmer, Patrick Maletinsky, Matthew Markham, Hongkun Park, Anders Sorensen, Daniel Twitchen, Alastair Stacey, Ron Walsworth, Jeorg Wrachtrup, and Amir Yacoby. Support for the work was provided by the NSF, CUA, DARPA, MURI, Packard Foundation, ERC Advanced Investigator Grant, and the NDSEG, NSF GRFP, Element Six, and Harvard Quantum Optics Center Fellowships.

\bibliographystyle{OUPnamed} 
\bibliography{NVChapter}

\end{document}